\documentstyle[graphicx]{mn2e}
\newif\ifAMStwofonts

\title{The contribution of star-forming galaxies to fluctuations in the 
cosmic background light.}
\author[Han-Seek Kim et al.]
       {Han-Seek~Kim$^{1,2}$\thanks{hansikk@unimelb.edu.au}, C. G.~Lacey$^1$, S.~Cole$^1$, C. M.~Baugh$^1$, C. S.~Frenk$^1$, G.~Efstathiou$^3$\\
       $^1$Institute for Computational Cosmology, Department of Physics, University of Durham, South Road, Durham DH1 3LE, UK\\
       $^2$School of Physics, University of Melbourne, Parkville, Victoria 3010, Australia.\\
       $^3$Kavli Institute for Cosmology Cambridge and Institute of Astronomy, Madingley Road, Cambridge, CB3 OHA, UK}
\date{}

\pagerange{\pageref{firstpage}--\pageref{lastpage}}
\pubyear{2010} 

\begin{document}

\maketitle
\title{The contribution of star-forming galaxies to fluctuations in the 
cosmic background light.}
\label{firstpage}

\begin{abstract}
Star-forming galaxies which are too faint to be detected individually 
produce intensity fluctuations in the cosmic background light. 
This contribution needs to be taken into 
account as a foreground when using the primordial signal to constrain 
cosmological parameters. The extragalactic fluctuations are also interesting 
in their own right as they depend on the star formation history of the Universe and 
the way in which this connects with the formation of cosmic structure. We present 
a new framework which allows us to predict the occupation of dark matter haloes by 
star-forming galaxies and uses this information, in conjunction with an N-body 
simulation of structure formation, to predict the power spectrum of intensity 
fluctuations in the infrared background. We compute the emission from galaxies 
at far-infrared, millimetre and radio wavelengths. Our method gives accurate 
predictions for the clustering of galaxies both 
for the one halo and two halo terms. We illustrate our new framework 
using a previously published model which reproduces the number counts and 
redshift distribution of galaxies selected by their emission at $850\,\mu$m. 
Without adjusting any of the model parameters, the predictions show encouraging 
agreement at high frequencies and on small angular scales 
with recent estimates of the extragalactic fluctuations in the background 
made from early data analysed by the Planck Collaboration. There are, however,  
substantial discrepancies between the model predictions and observations 
on large angular scales and at low frequencies, which illustrates the usefulness 
of the intensity fluctuations as a constraint on galaxy formation models.
\end{abstract}

\begin{keywords}
galaxy clustering, Planck
\end{keywords}

\section{Introduction}

The cosmic background light (CBL) is a rich source of information 
about the conditions in the early universe and the 
subsequent growth of galaxies and of structure in the dark matter. Accurate 
measurements of the power spectrum of temperature anisotropies in the primordial 
component have led to tight constraints being placed on the values of the basic cosmological 
parameters (e.g. Komatsu et~al. 2011). Other contributions to the CBL include Galactic cirrus, 
the thermal and kinetic Sunyaev-Zel'dovich effects and extragalactic 
sources such as star-forming galaxies (SFGs) and active radio galaxies 
 (see e.g. Figure 2 of Dunkley et al. 2011). 
{\bf Correlated} fluctuations in the CBL due to SFGs depend on the number of sources 
and their clustering, which in turn is sensitive to the variation 
in the efficiency of star formation with dark matter halo mass. The 
extragalactic contribution to the CBL may be viewed either as a nuisance 
to be removed statistically in order to get to the primordial CBL signal, 
or as an interesting quantity in its own right as a challenge to models of 
the clustering of galaxies and their emission in the infra-red, millimetre 
and radio ranges of the electromagnetic spectrum.  

A small fraction, around 10-20\%, of the extragalactic background light  
has been resolved into galaxies at far-infrared and millimetre 
wavelengths (Bethermin et~al. 2010; Oliver et~al. 2011). 
Fluctuations in the intensity of the unresolved CBL have been discovered 
recently (Grossan \& Smoot 2007; Lagache et al. 2007; Viero et al. 2009;  
Hall et al. 2010; the Planck Collaboration 2011b; Penin et~al. 2012a). 
These fluctuations have 
two sources: the shot noise arising from sampling a discrete 
number of unresolved galaxies within the telescope beam and the 
intrinsic clustering of the galaxies. In an early analysis of 
six regions of low Galactic extinction covering $140$ square 
degrees, the Planck Collaboration (2011b) 
have cleaned the temperature anisotropy maps to leave only 
the extragalactic fluctuations in the CBL. This is done by 
using the lowest frequency Planck map to remove the cosmological 
signal (we shall see later that fluctuations due to the clustering of extragalactic 
sources are negligible at low frequencies) and exploiting neutral hydrogen  
observations as a tracer of dust to further reduce the contribution from 
Galactic emission. (For an overview of the Planck mission, see the 
Planck Collaboration 2011a.) 

A variety of models have been developed to interpret the measured 
power spectrum of fluctuations in the intensity of the CBL. These 
models have predominantly used empirical spectral energy distributions 
(e.g. Lagache et~al. 2003, 2007). Simple analytic models have been 
assumed for the clustering of galaxies such as a linear bias factor 
relative to the clustering of the dark matter (e.g. Knox et~al. 2001; 
Fernandez-Conde et~al. 2008; Hall et al. 2010) or the halo occupation 
distribution (HOD) formalism (Amblard \& Cooray 2007; Viero et al. 2009; 
Shang et~al. 2011; Penin~et al. 2012b). 
Righi et~al. (2008) presented a calculation based on the 
mergers of dark matter haloes and a simple dust evolution model.
Seghal et al. (2010) assumed that the number of infra-red sources 
scaled with halo mass to compute the two halo clustering, without calculating 
a one halo contribution. 
Negrello et~al. (2007) combined the model of Granato et~al. (2004) for 
the evolution of the spheroid population with a phenomenological model 
for the evolution of starbursts, normal late-type spirals and radio galaxies. 
These authors assumed a linear bias to model the clustering of galaxies. 
The Planck Collaboration (2011b) have used their measurements of the CBL 
fluctuations to rule out a linear bias model that is constrained to 
match the observed number counts of galaxies, arguing that accurate  
small-scale clustering predictions are critical to match the observations. 

In this paper we present a new approach for computing the contribution of SFGs 
to the intensity fluctuations in the CBL, with two important 
improvements over previous theoretical models. First, we make 
an ab initio calculation of the spectral energy distributions 
of a large sample of galaxies, using a self-consistent 
treatment of the extinction of starlight by dust and 
the reprocessing of the absorbed energy to longer wavelengths. 
We also compute the radio emission from star-forming galaxies 
(Condon 1992; Bressan, Silva \& Granato 2002). Second, we combine the 
predictions for the properties of the galaxy population with a high resolution, 
large volume N-body simulation of the clustering of matter in 
the Universe. This allows us to accurately model the clustering of 
galaxies over a wide range of pair separations, including the highly nonlinear 
regime corresponding to scales within an individual dark matter halo.   
Empirical models suffer from the obvious drawback that the bulk 
of the galaxies responsible for the extragalactic background have not 
yet been observed, which makes the calibration of this class of model 
uncertain. Also, without the context of a model for structure 
formation, any assumptions about the clustering of the galaxies are 
decoupled from their abundance (e.g. as in the calculation by Xia et~al. 2012). 

The first step in our calculation is to generate predictions for the 
star formation histories of a representative sample of galaxies, using 
the semi-analytical galaxy formation code, {\tt GALFORM} (Cole et~al. 
2000; Baugh et~al. 2005). The star formation and metal enrichment 
histories, along with the size of the disc and bulge components 
are input to the spectro-photometric code {\tt GRASIL} (Silva et~al. 1998). 
{\tt GRASIL} is used to compute the spectral energy distribution of each galaxy, 
using a radiative transfer calculation in a two-phase dust medium (Granato 
et~al. 2000).
We describe how the hybrid {\tt GALFORM} plus {\tt GRASIL} code is 
implanted into a large-volume, high resolution N-body simulation of 
the clustering of the dark matter to add information 
about the spatial distribution of galaxies. This allows an 
accurate treatment of the one-halo term and of the nonlinear component 
of the two-halo term.

The content of the paper is as follows. In Section~2 we give a brief 
overview of the {\tt GALFORM} and {\tt GRASIL} models before explaining 
how we populate an N-body simulation with model galaxies. In Section~2.3 
we set out the calculation of the angular correlation function of flux. 
The results of the paper are presented in Section~3 and the summary in 
Section~4. The appendix discusses the sensitivity of the model 
predictions to the finite resolution of the N-body simulation.

\section{Theoretical background} \label{modell}

In this section we introduce the galaxy formation model used and 
outline the theoretical concepts needed in the paper. We give a 
brief overview of the semi-analytical galaxy formation model  
and explain how the emission from dust and at radio wavelengths 
is computed in Section~\ref{model}. 
The implementation of this model in an N-body simulation is described 
in Section~\ref{GTON}. In Section~\ref{MOACF} we set out the equations describing 
how the clustering of intensity fluctuations due to extragalactic 
sources is computed from the predictions of the galaxy formation model.

\begin{table*}
\caption{
The frequencies covered by the Planck instruments.
The {\it rows} give: (1) The name of the instrument.
The first three frequencies (columns 2-4) correspond to 
the Low Frequency Instrument (LFI) and columns 5-10 to the 
High Frequency Instrument (HFI). 
(2) The central frequency of the channel in gigahertz. 
(3) The central wavelength in millimetres. 
(4) The flux limit for point sources in Janskys (taken from Vielva et~al. 2003 in the 
case of LFI and the Planck Collaboration 2011b for the HFI). 
(5) The angular resolution in arcminutes.} 
\label{PlanckIn}
\begin{tabular}{lccccccccc}
\hline
 & LFI & LFI & LFI & HFI & HFI & HFI & HFI & HFI & HFI \\
\hline
\hline
Frequency (GHz) & 30& 44& 70& 100&143&217&353&545&857\\
Wavelength (mm) & 10& 6.81& 4.29& 3.0&2.1&1.38&0.85&0.55&0.35\\
Flux limit (Jy) & 0.23& 0.25& 0.24& 0.25&0.25&0.16&0.33&0.54&0.71\\
Angular resolution (arcmin) & 33& 24& 14& 9.5&7.1&5.0&5.0&5.0&5.0\\
\hline
\end{tabular}
\end{table*}

\subsection{The hybrid galaxy formation model}\label{model}

We use a hybrid model consisting of the {\tt GALFORM} semi-analytical 
galaxy formation code (Sec~\ref{GFM}) and  
the {\tt GRASIL} spectrophotometric code (Sec~\ref{GRM}).

\subsubsection{The GALFORM galaxy formation model}
\label{GFM}

The formation and evolution of galaxies within the $\Lambda$CDM cosmology 
is predicted using the semi-analytical model {\tt GALFORM} (Cole et~al. 2000). 
The main processes modelled include: (1) the formation of dark matter 
haloes by mergers and the accretion of smaller haloes, (2) the growth 
of galactic discs following the shock-heating and radiative cooling of gas 
inside dark matter haloes, (3) star formation in galactic discs, 
(4) the reheating and ejection of gas by supernovae,
(5) the prevention of gas cooling in low circular velocity haloes due to 
the photoionization of the intergalactic medium, (6) the loss of galaxy 
orbital energy through dynamical friction, (7) the subsequent 
merger between galaxies, which may be accompanied by a burst of 
star formation and (8) chemical evolution of the stars and gas.
By following these processes, {\tt GALFORM} predicts the star formation 
history of each galaxy (see Baugh 2006 for a review). 
The code can be run quickly for a representative sample of 
dark matter haloes, making it ideal for generating predictions 
for number counts and to populate large volumes to predict galaxy clustering. 

The galaxy formation model we use in this paper is that of Baugh et~al. (2005; 
see also Lacey~et al. 2010). This model reproduces the observed number counts 
and redshift distribution of galaxies at $850\,\mu$m, and also the luminosity 
function of Lyman break galaxies (see Lacey et~al. 2011). 
(Note that the more recent model of Bower et~al. 2006 does not enjoy these successes, 
and so is not used in this paper.) 
The background cosmology is a spatially flat $\Lambda$CDM model 
(see later for the values of the cosmological parameters).
The merger histories of the dark matter haloes are generated using an  
improved Monte Carlo technique that has been calibrated against 
merger trees extracted from an N-body simulation (Parkinson, Helly \& Cole 2008).

The model follows two modes of star formation, a ``quiescent" mode 
which takes place in galactic discs and a ``burst" mode which is 
triggered by major and minor galaxy mergers. 
Mergers are classified according to the ratio of the mass of the 
merging satellite, ${\rm M_{sat}}$, to that of the central galaxy, 
${\rm M_{cen}}$. If ${\rm M_{sat}/M_{cen}} 
\ge f_{\rm ellip}$ (where $f_{\rm ellip}$ is a model parameter) 
then the merger is defined as a major merger. In this case, 
any cold gas in the two galaxies takes part in a starburst which adds stars 
to the spheroid. Minor mergers which have mass ratios in the range 
$f_{\rm burst} < M_{\rm sat}/M_{\rm cen} < f_{\rm ellip} $ and where 
the primary is also gas rich, with 
$M_{\rm cold}/ \left( M_{*} + M_{\rm cold} \right) > f_{\rm gas}$ 
are also assumed to trigger starbursts. In the Baugh et~al. model 
the parameter values 
were set to $f_{\rm ellip} = 0.3$, $f_{\rm burst} = 0.05$ and 
$f_{\rm gas} = 0.8$. Different stellar initial mass functions (IMF) 
are adopted in the two modes of star formation. Quiescent star formation 
is assumed to take place with a solar neighbourhood IMF (Kennicutt 1983). 
Bursts of star formation are assumed to form stars with a top heavy IMF. 
Baugh~et al. (2005) argued that the adoption of a top heavy IMF in bursts 
is necessary to reproduce the observed number counts and redshift 
distributions of the faint sub-mm galaxies, whilst at the same time 
reproducing the properties of the local galaxy population. 

\subsubsection{The GRASIL spectrophotometric code}
\label{GRM}

The frequencies sampled by the Planck Low Frequency Instrument 
(LFI; the Planck Collaboration 2011c) and High Frequency Instrument 
(HFI; the Planck Collaboration 2011d) are listed in Table~\ref{PlanckIn}. 
At these frequencies (corresponding to wavelengths $0.3$-$10$mm) 
we assume that the contribution from galaxies is 
dominated by star-forming galaxies through dust heated by stars and radio 
emission resulting from gas ionized by stars and synchrotron radiation from 
relativistic electrons accelerated in supernova remnant shockwaves. 
We do not attempt to model the contribution to the CBL of active 
galactic nuclei or galaxies in which the radio emission is powered 
by accretion onto a central black hole.  
To predict the emission from model galaxies 
at these frequencies we use the {\tt GRASIL} spectrophotometric code (Silva et~al. 1998; 
Bressan et~al. 2002). 
{\tt GRASIL} computes the emission from the composite stellar population 
of the galaxy, using theoretical models of stellar evolution and 
stellar atmospheres. The interaction of the starlight with dust is 
followed with a radiative transfer calculation which assumes a 
two-phase dust medium, and gives the distribution of dust temperatures 
within each galaxy using a detailed grain model.  
The output from {\tt GRASIL} is the galaxy SED from the far-UV 
to the radio (wavelengths 0.01\,$\mu$m $\leq$ $\lambda$ $\leq 1\, {\rm m}$). 
The main features of the hybrid {\tt GALFORM-GRASIL} model are described in 
Lacey et al. (2010; see also Granato et~al. 2000). 

\subsection{Populating an N-body simulation with galaxies}\label{GTON}

\begin{figure}
\includegraphics[width=8.6cm]{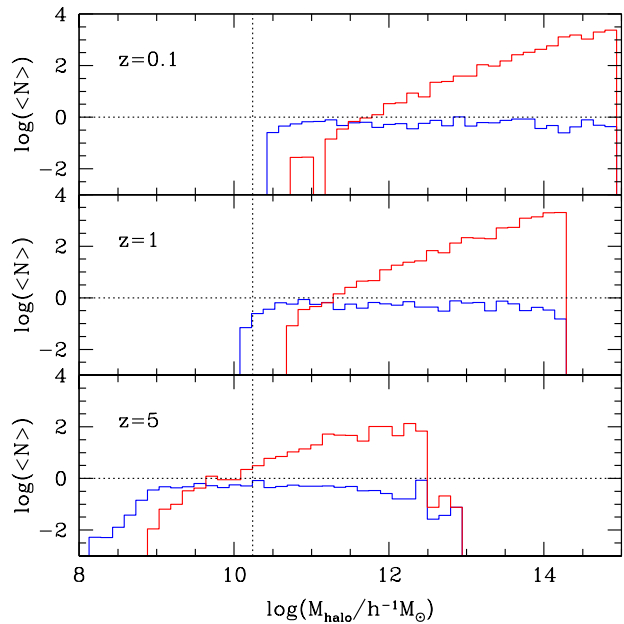}
\caption{
The halo occupation distribution (HOD) of galaxies in the MCGAL 
catalogue at different redshifts, as indicated by the labels. 
The blue lines show the HOD of central galaxies and 
red lines show the HOD of satellite galaxies. The vertical line 
indicates the halo mass resolution of the Millennium simulation 
and the horizontal line shows $\langle N \rangle =1$. All galaxies 
with stellar mass in excess of $10^{7} h^{-1} M_{\odot}$ are allowed 
to contribute to the HOD plotted. This is the input HOD used to build 
the MILLGAL catalogue. 
}
\label{HODGRASIL}
\end{figure}

\begin{figure}
\includegraphics[width=8.6cm]{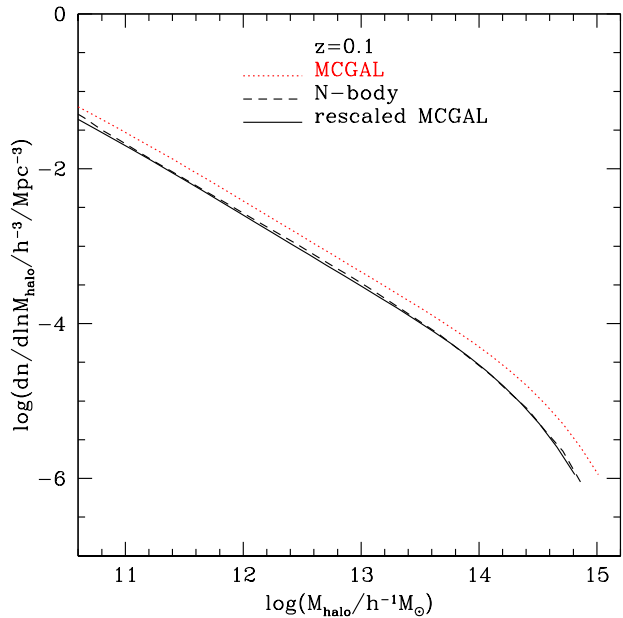}
\caption{
The dark matter halo mass function at $z=0.1$. The red dotted line shows the 
halo mass function used in the MCGAL catalogue. The black dashed line 
shows the halo mass function in the Millennium simulation at this 
redshift. The black solid line shows the rescaled MCGAL halo mass function,  
after applying a single mass independent adjustment to the halo mass in 
the MCGAL cosmology. 
} 
\label{MASSFC}
\end{figure}

We combine the hybrid {\tt GALFORM-GRASIL} galaxy formation model with a high 
resolution N-body simulation of the clustering of matter. This 
allows us to make accurate predictions for the clustering of galaxies 
selected by their emission at infrared, millimetre and radio wavelengths. In particular the 
small-scale clustering measured in the simulation can be significantly 
different in practice from simple analytical expectations based on linear perturbation theory 
(as shown, for example, by Benson et~al. 2000). We shall see later that for some  
frequencies, the clustering of galaxy pairs within the same dark matter 
halo is important for the power spectrum of the microwave background 
intensity fluctuations on the angular scales probed by Planck. 

We now describe how galaxies are implanted into 
an N-body simulation. 
Our starting point is a set of model galaxies generated using 
the hybrid {\tt GALFORM} plus {\tt GRASIL} code set in the 
concordance $\Lambda$CDM cosmology, which we refer to as the MCGAL catalogue. 
The end point is a galaxy catalogue implanted in the Millennium Simulation of 
Springel et~al. (2005), an N-body simulation of a cosmologically  
representative volume.
We denote this as the MILLGAL catalogue. 
Some minor complications arise because 
the cosmology adopted in our original calculation is not quite 
the same as that of the N-body simulation; this issue is dealt 
with in step 2 below. The most accurate calculation of galaxy clustering 
is made using the {\tt MILLGAL} catalogue, so we focus on this model 
in the main paper. In some instances we show predictions made using 
both versions of the model to illustrate regimes in which the {\tt MILLGAL} 
predictions are superior. The {\tt MILLGAL} calculation has a finite 
resolution; as we show in the Appendix, this does not affect our results.

The MCGAL catalogue is constructed by sampling galaxies according to their 
stellar mass from a much larger catalogue of galaxies. 
This larger catalogue is generated from halo merger histories generated using a 
Monte Carlo method (Cole et~al. 2000; Parkinson et~al. 2008). 
Star formation histories are extracted for the selected galaxies, which are then input to {\tt GRASIL}, 
along with the predicted dust masses and sizes of the disc and bulge components,  
to compute the galaxy SED. For each galaxy we have the stellar mass, host 
halo mass, a weight based on the halo number density and on the sampling rate as 
a function of stellar mass, and the galaxy SED. Due to the computational 
expense of generating the {\tt MCGAL} catalogue, and its similarity with the 
{\tt MILLGAL} catalogue which we wish to build, we have devised the following 
scheme to take advantage of the availability of this calculation, rather than 
making a new calculation based on star formation histories extracted directly 
from galaxies in the Millennium simulation.

The steps followed to populate the Millennium simulation with galaxies 
starting from the MCGAL catalogue are as follows: 

\begin{itemize}
\item[(1)] 
{\it Construct the halo occupation distribution (HOD) of 
galaxies.} 
{\tt GALFORM} predicts how many galaxies are contained within each 
dark matter halo. The HOD quantifies the mean number of galaxies per halo as a 
function of the halo mass (Benson et~al. 2000; Peacock \& Smith 2000; 
Berlind \& Weinberg 2002). Fig.~\ref{HODGRASIL} shows the HOD of galaxies 
in the MCGAL catalogue at three different redshifts. We plot the HOD 
for central (blue line) and satellite (red line) galaxies separately. 
Fig.~\ref{HODGRASIL} shows that the HODs of central galaxies in 
the MCGAL catalogue are approximately step functions. The shape of the 
satellite galaxy HOD is close to a power law in halo mass. 
The lowest mass dark matter halo which appears in the MCGAL HODs 
varies with redshift. The halo mass grid used in this calculation is 
defined at each redshift in order to sample haloes with a representative 
range of abundances. At $z=0.1$, the lowest mass halo 
considered is nearly the same as the lowest mass halo which can be resolved 
in the Millennium simulation ($\sim 10^{10.3}h^{-1} {\rm M}_{\odot}$, which 
is shown by the vertical dotted line in Fig.~\ref{HODGRASIL}). The lowest 
mass dark matter halo in the Millennium simulation is the same at 
all redshifts. We cannot transplant galaxies from 
the MCGAL catalogue which reside in haloes below the mass resolution 
of the Millennium. We test the sensitivity of our predictions 
to this limitation of the N-body catalogue in the Appendix. 

\item[(2)] {\it Match the halo mass function between the cosmologies used.} 
The MCGAL catalogue, for historical reasons, 
assumes a slightly different cosmology to that 
adopted in the Millennium simulation. The 
Millennium cosmology is based on the first year of WMAP observations\footnote{The cosmological parameters in the Millennium are: matter density $\Omega_{\rm M}=0.25$, cosmological constant $\Omega_{\Lambda}=0.75$, Hubble constant 
$H_{0}=73 \,{\rm km \,s}^{-1}\,{\rm Mpc}^{-1}$, primordial scalar 
spectral index $n_{\rm s}=1$, baryon density $\Omega_{\rm b}=0.045$ 
and fluctuation amplitude $\sigma_{8}=0.9$. The parameters 
in the MCGAL case are $\Omega_{\rm M}=0.3$, $\Omega_{\Lambda}=0.7$, 
$H_{0}= 73 \,{\rm km \,s}^{-1}\,{\rm Mpc}^{-1}$, $n_{\rm s}=1$, 
$\Omega_{\rm b} = 0.04$, and $\sigma_{8} = 0.93$ (Baugh et~al. 2005).}. 
At a given mass, the number density of haloes is slightly different 
in the two cosmologies. 
The {\tt GALFORM-GRASIL} code is computationally expensive to run, which prohibits 
running the calculation again in the Millennium cosmology. A re-run would also 
require some retuning of the galaxy formation parameters. To transplant 
the galaxies from the halo population in one cosmology to that in the other 
cosmology, we instead chose to relabel the halo masses in the 
MCGAL model to force a match with the Millennium simulation mass function.
Fig.~\ref{MASSFC} shows the halo mass functions in the original MCGAL calculation 
(red dotted line) and in the N-body simulation (black dashed line), 
along with the rescaled MCGAL halo mass function (black solid line) at $z=0.1$. 
To match the halo mass functions at $z=0.1$ we reduce the halo mass globally 
in the MCGAL catalogue by a factor of $1.2$. The halo mass function of 
the rescaled MCGAL catalogue and that of Millennium simulation 
agree to better than 5\% over four decades in halo mass. 
We apply the same scheme to other redshifts and find that it works equally well, 
but with slightly different scaling factors.
The small difference between the halo mass functions in the two cosmologies 
supports our decision not to rerun the {\tt GALFORM-GRASIL} calculation in the 
cosmology of the Millennium simulation.

\item[(3)] {\it Place galaxies in the N-body simulation.} 
We generate the MILLGAL catalogue of galaxies using the MCGAL HOD with  
halo masses rescaled, as explained in Step 2. The central galaxy is placed at the 
centre of mass of the halo. In the case of halo masses for which the 
HOD specifies $N<1$, a fraction of the haloes of this mass are populated with a single 
galaxy at random with a probability $N$. 
The number of satellite galaxies in a given halo 
mass is assumed to have a Poisson distribution for $N>1$, with the mean number of 
satellite galaxies given by the HOD. 
Satellite galaxies are assigned to randomly selected dark matter particles which are part 
of the friends-of-friends halo. 
\end{itemize}

\subsection{Calculation of the angular power spectrum of 
intensity fluctuations}
\label{MOACF}

In this subsection we give the theoretical background to the 
calculation of the angular power spectrum of the intensity 
fluctuations of galaxies. The discussion is split into three parts: 
\S2.3.1 The calculation of the angular correlation function of intensity 
fluctuations. 
\S2.3.2 The calculation of the angular power spectrum of intensity 
fluctuations. 
\S2.3.3 The estimation of the spatial correlation function 
of intensity fluctuations. 
The first two parts are completely general. In the final section 
we outline how the luminosity-weighted spatial correlation function 
is estimated in the two cases we consider: the direct, simulation based 
approach (MILLGAL) and the analytical calculation (MCGAL). Throughout 
we discuss various quantities which depend on intensity at 
a particular frequency. For ease of reading, we suppress the explicit 
frequency dependence in our notation. For example, we write the 
luminosity density at position $x$, $\rho_{L}(x,\nu)$ as $\rho_{L}(x)$.
We remind the reader that all quantities which depend on intensity or 
luminosity have a frequency dependence. 

\subsubsection{Calculation of the angular correlation function 
of intensity fluctuations}

We can define the spatial correlation function of luminosity density,
$\xi_{L}(\vec{x})$, as
\begin{equation}
\langle \rho_{L}(\vec{x}_1) \rho_{L}(\vec{x}_2) \rangle = \langle \rho_{L} \rangle^2 \ \left(
1+\xi_{L}(\vec{x}_1-\vec{x}_2)\right),
\end{equation}
where $\rho_{L}(\vec{x}_1)$ is the luminosity density at position $\vec{x}_1$ 
and $\langle \rho_{L} \rangle$ is the mean luminosity density.
Note that here we neglect shot noise (see \S2.3.2) 
by assuming $x_{1} \ne x_{2}$. 
The correlation function of galaxy luminosity is related to the 
standard spatial correlation function through
\begin{equation}
\xi_{L} (r) = \frac{ \int \xi(L_{1},L_{2},r) L_1 L_2 n(L_1) n(L_2) 
{\rm d} L_1 {\rm d} L_2 }{ \left( \int n(L) {\rm d} L \right)^2 }, 
\end{equation} 
where $n(L)$ is the mean number density per unit luminosity 
of objects with luminosity $L$, $L_{i}$ is the luminosity of 
the $i^{\rm th}$ galaxy in the pair, 
and $\xi(L_{1},L_{2},r)$ is the cross-correlation function of galaxies 
with luminosities $L_1$ and $L_2$. 
We can write the luminosity density as 
$\langle \rho_{L} \rangle = \int n(L) {\rm d} L$. 
Similarly, we can define an angular surface brightness
correlation function, $w_{\rm I}(\vec{\theta})$, as
\begin{equation}\label{fv}
\langle I(\vec{\theta_1})I(\vec{\theta_2})\rangle= 
\langle I \rangle^2 \ \left(
1 + w_{\rm I}(\vec{\theta_1}-\vec{\theta_2})\right) ,
\end{equation}
where $\theta$ is in radians and the surface brightness 
is related to the comoving luminosity density via
\begin{equation}
  I(\vec{\theta})\, {\rm d}^2\theta = \frac{1}{4\pi} \, \int \frac{x^2 \rho_{L}}{{\rm d}_{L}^2(x)} 
\, {\rm d}x \, {\rm d}^2\theta,
\end{equation}
where we have assumed a geometrically flat universe ($\Omega_{\rm total} = 1$), and 
hence $d_{L}(x)$, the luminosity distance to comoving distance $x$, 
is given by $d_{L}(x) = (1 +z )x$.
The above equation applies in the case of bolometric luminosity
densities and intensities. However, in practice we are nearly always interested in
fluxes that are measured over a limited frequency band. This introduces an 
extra $(1+z)$ factor to account for the change in the band width with redshift, 
giving an intensity per unit frequency of 
\begin{equation}
  I(\vec{\theta})\, {\rm d}^2\theta = \frac{1}{4\pi} \, \int \frac{(1+z) x^2 
\rho_{L}}{{ d}_{L}^2(x)} 
\, {\rm d}x \, {\rm d}^2\theta,
\end{equation}
where, if $\nu$ is the observed frequency, then $\rho_{L}$ is now the 
luminosity density per unit frequency in a band centred on the 
rest-frame frequency $\nu(1+z)$. (NB the frequency dependence 
of the intensity is suppressed in our notation.) 
With the assumption of a spatially flat universe, the mean intensity is 
given by:  
\begin{equation}
\langle I \rangle = \frac{1}{4\pi} \, 
\int_0^\infty \frac{\langle \rho_{L }(x)\rangle}{(1+z)} \, {\rm d}x .
\label{FZEQ}
\end{equation}

We use Limber's approximation to obtain an expression for the angular clustering of 
flux from the spatial correlation function (Limber 1953). 
First, it is assumed that the mean number density of galaxies, 
$\langle n(x) \rangle$, 
varies sufficiently slowly with redshift (here labelled by the comoving radial
coordinate $x$) that over the range of pair separations for 
which $\xi(\vec{x}_1-\vec{x}_2)\ne0$,
$\langle n(x_1) \rangle \approx \langle n(x_2)\rangle$. Second, we assume the small 
angle approximation, i.e. the angular separation of pairs of galaxies for which
$\xi(\vec{x}_1-\vec{x}_2)\ne0$ is small (i.e. $ | \vec{\theta_{1}} - \vec{\theta_{2}} | \ll 1$ with 
$\theta$ in radians). 

Using the above approximations, we can relate the spatial correlation function 
of galaxies, $\xi(r)$, to the angular correlation, $w(\theta)$, through Limber's equation
\begin{equation}
w(\theta) 
= \frac{
\int_0^\infty x^4  \,  \int_{-\infty}^\infty 
\langle n(x) \rangle^2\xi((u^2 + x^2\theta^2)^{1/2})\, {\rm d}x\, {\rm d}u }{\left(\int x^2 n(x) {\rm d}x \right)^2} , 
\end{equation}
where $u$ is a comoving separation parallel to the line of sight, such 
that $r^{2} = u^{2} + x^{2} \theta^{2}$ 
(again, for small angle separations).  
The analogous relation to Limber's equation for $w_{{\rm I}}(\theta)$ 
is given by  
\begin{eqnarray}
&& \hspace{-0.7cm} w_{{\rm I} }(\theta) = \left(\frac{1}{4\pi}\right)^2 \, \frac{1}{\langle I \rangle^2} \\
&& \hspace{-0.7cm} \int_{-\infty}^{\infty} \int_0^{\infty}
\frac{(\langle \rho_{L}(x)\rangle^2 }{ (1+ z)^2} \ \xi_{L}\left(
(u^2+x^2\theta^2)^{1/2}
 \right)
{\rm d}x \, {\rm d}u . \nonumber 
\end{eqnarray}

The correlation function of intensity can be evaluated at discrete 
redshifts to give
$\langle \rho_{L}(x)\rangle$ and $\xi_{L}(r,x)$ (where redshift 
is again labelled by radial comoving distance $x$) which can then be 
input into Eq.~8 to compute the angular clustering of the flux.
Note that in the calculations presented in this paper we are 
interested in the galaxies with flux fainter than the detection 
limits in the Planck bands, as listed in Table 1. We test the impact of the 
finite resolution of the N-body sample on our predictions in the 
Appendix. 
The quantities we need to calculate are $\langle I \rangle$ at
the frequencies corresponding to the Planck bands and the fluctuations
in this background, which are given by $ \langle I \rangle^2\, w_{{\rm I}}(\theta)$.
These quantities are predicted by the galaxy formation model described 
in the previous subsections.

\subsubsection{Calculation of the angular power spectrum}

The angular power spectrum of the intensity fluctuations can 
be obtained from the angular correlation function of intensity using 
\begin{equation}
C_{l }(\theta)= 2\,\pi \langle I \rangle^{2} \int^{\pi}_{0} 
w_{I }
(\theta)P_{l}(\cos\theta)\sin\theta{\rm d}\theta.
\end{equation}

The discrete nature of the sources contributing to the CBL, even 
though they may not be resolved individually by an instrument such as Planck, 
leads to shot noise in the intensity fluctuation power spectrum. Even if the 
galaxies displayed no intrinsic clustering and were distributed at random, they 
would make a contribution to the power spectrum through the shot noise. 
The contribution to the power spectrum from the shot noise depends on the 
number of sources (Tegmark \& Efstathiou 1996): 

\begin{equation} 
C_{l} = \int_{0}^{S^{\rm lim}} S^{2} \frac{ {\rm d} N}{ {\rm d} S} 
{\rm d} S ,
\label{shot}
\end{equation} 
where $S^{\rm lim}$ is the flux detection limit in a given band.

\subsubsection{Estimation of the intensity-weighted correlation function}

We follow two different approaches to estimate the spatial 
clustering of galaxies, depending on whether we are using the 
MILLGAL catalogue (for which we have galaxy positions) or the 
MCGAL catalogue (for which we know the mass of the halo which 
hosts each galaxy). In both cases, there is an assumption that the 
clustering of haloes in which galaxies are placed depends only 
on their mass and not on their environment (for an assessment 
of how halo clustering depends on properties besides mass 
see e.g. Gao et~al. 2005; Angulo et~al. 2008). 

The primary method is to compute the spatial luminosity-weighted 
galaxy correlation function directly, using the galaxies 
transplanted into the N-body simulation (i.e. the MILLGAL catalogue). 
The correlation function is estimated from the pair counts 
of galaxies.\footnote{The simulation volume is periodic so 
the volume of the spherical shell for pair separations in the range $r$ 
to $r+ {\rm d}r$ can be calculated analytically; see e.g. Eke 
et~al. 1996 for the form of the estimator for the two-point 
correlation function in this case.} This method naturally 
accounts for any difference between the clustering of the galaxies 
and the underlying dark matter because of the imposition of the HOD. 
This approach also allows an accurate prediction of the correlation 
function on small scales, corresponding to galaxy pairs within the 
same dark matter halo. 

The second approach (used in conjunction with the MCGAL catalogue) 
is analytical and is intended to show on which scales 
the improvement comes from using the N-body simulation. 
In this case there are two steps in the calculation. The first 
is step is to compute the effective clustering 
bias of the galaxy sample, using the HOD to perform 
a weighted average of the bias of each galaxy based on the analytical 
halo bias (Sheth, Mo \& Tormen 2001; see Kim et~al. 2009 for the steps 
connecting the HOD to the effective bias). 
The analytic estimate of the clustering accounts for only the 
2-halo term, and assumes that at large separations, the bias 
is independent of scale and depends only on halo mass. 
The effective bias of a galaxy sample is given by
\begin{equation}
b_{L, {\rm eff}} = \frac{ 
\int \int b(M) L N (L | M) n(M) {\rm d} L {\rm d} M }
{\int \int L N(L|M) n(M) {\rm d} L {\rm d} M }, 
\end{equation}
where $b(M)$ is the clustering bias factor for haloes of mass $M$, 
$N(L|M)$ is the HOD (the mean number of galaxies per halo which 
satisty the sample selection i.e. have luminosity $L$) as a function 
of halo mass $M$ and $n(M)$ is the halo mass function, which gives 
the abundance per unit volume per $d \ln M$ bin of haloes of mass $M$. 

The second step is to generate a correlation function 
for the dark matter. This is done by 
Fourier transforming the nonlinear power spectrum of the mass derived using 
the approximate analytical prescription of Smith et~al. (2003). The galaxy correlation function is 
then obtained by multiplying the nonlinear matter correlation function by 
the square of the effective bias.


\begin{figure*}
\includegraphics[width=8.6cm]{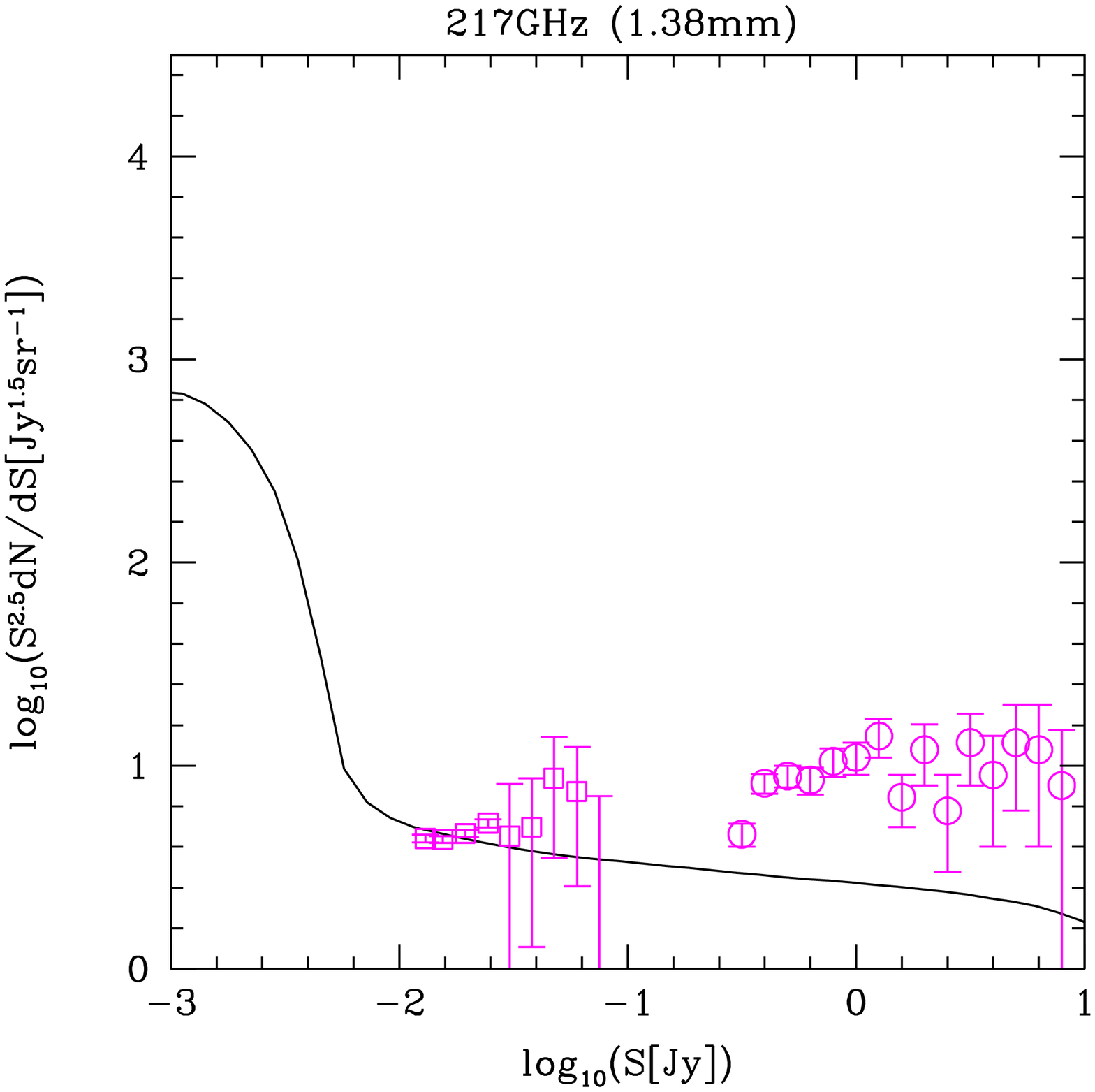}
\includegraphics[width=8.6cm]{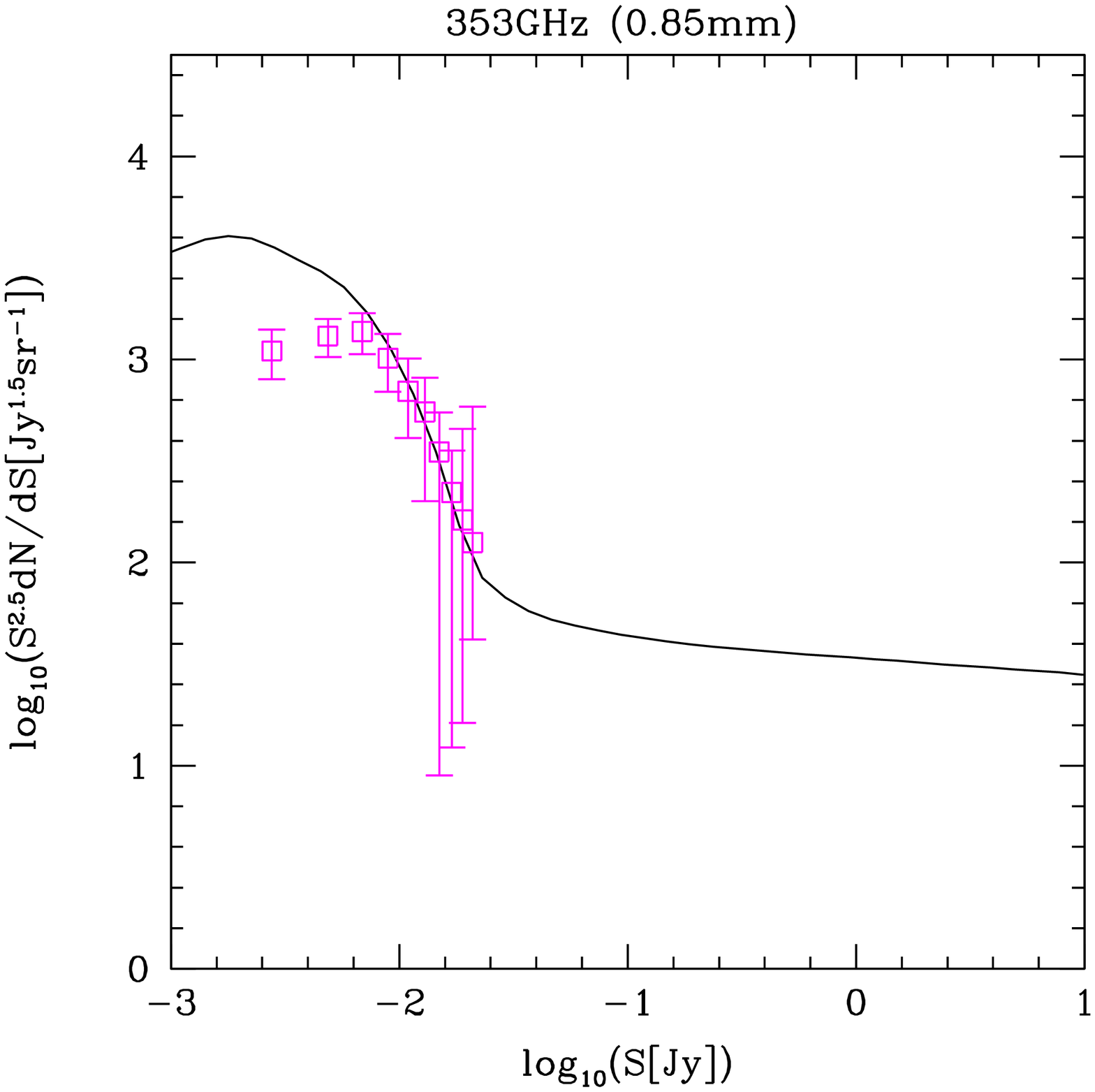}
\includegraphics[width=8.6cm]{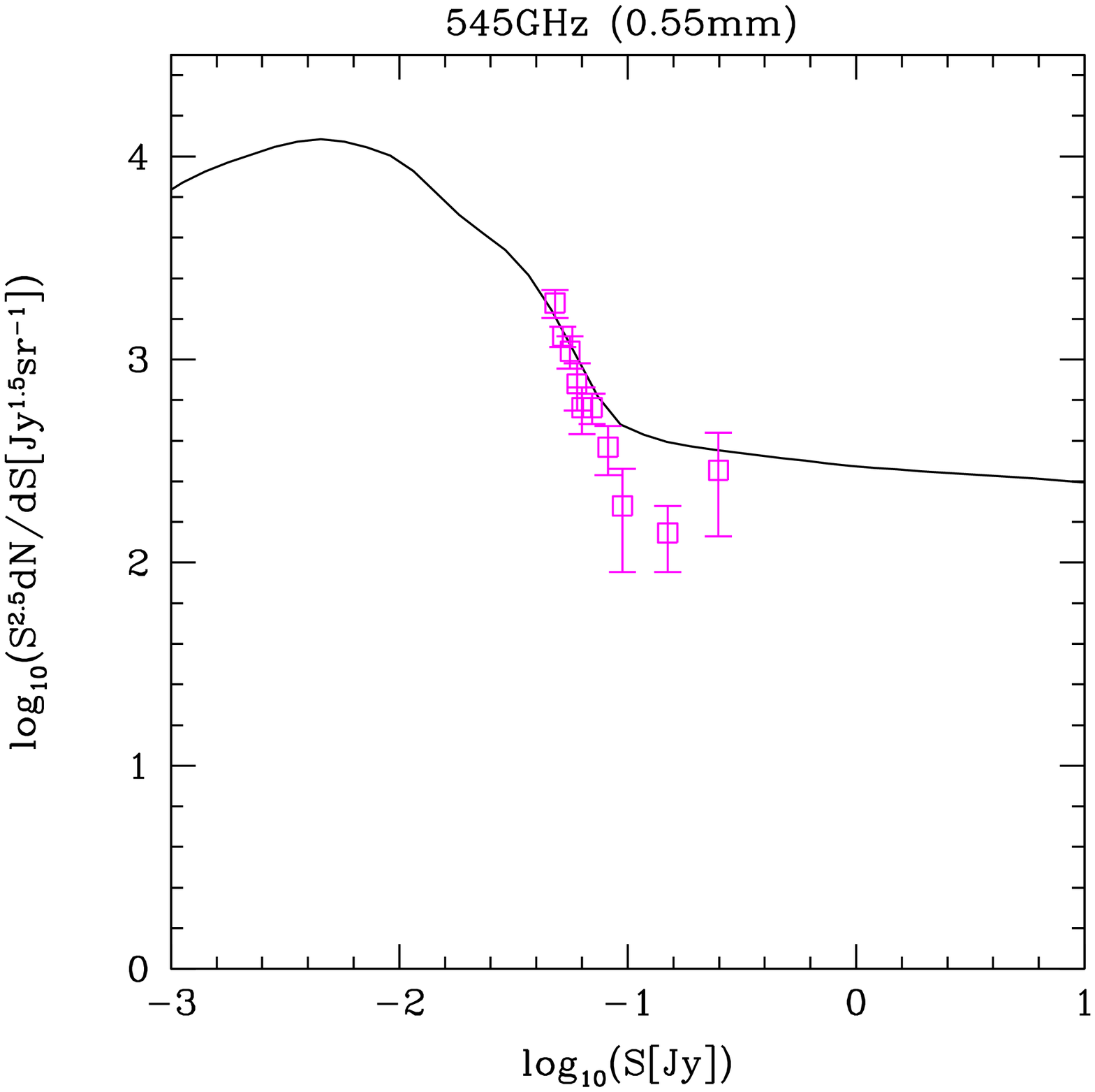}
\includegraphics[width=8.6cm]{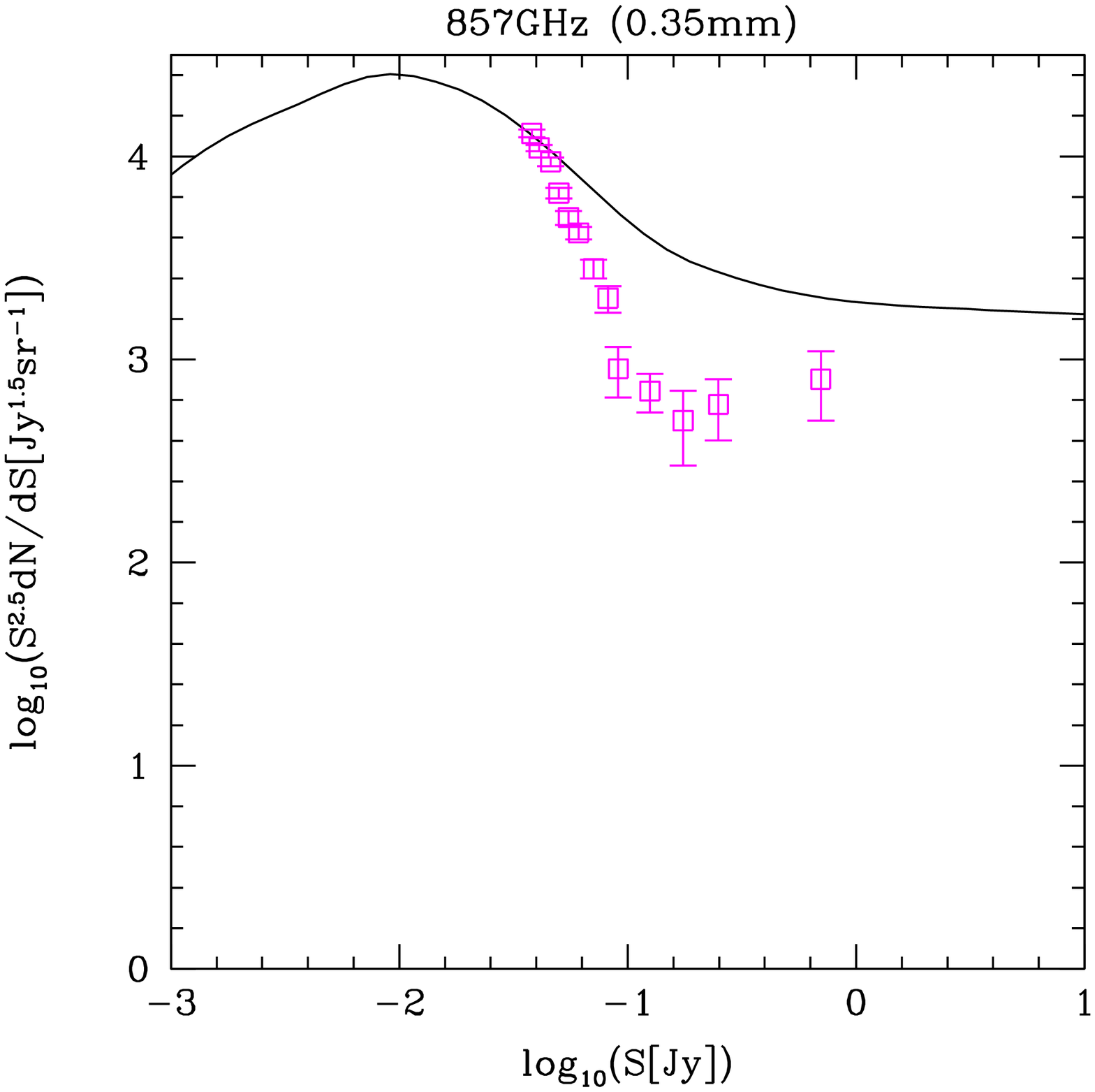}
\caption{
The predicted differential number counts from the MCGAL model (solid lines) 
compared with observational counts (points) in the Planck HFI bands, in 
some cases measured by other telescopes at similar wavelengths. 
Note that we have adopted the cosmology of the MILLGAL model, rescaling the 
halo masses as described in Sec.~2.2 and using the abundance of haloes in 
the MILLGAL cosmology. The observational data comes from: 
at $217\,$GHz, the South Pole Telescope (Vieira et~al. 2010; squares) and 
the Planck Collaboration 2011e (circles); 
at $353\,$GHz  from Coppin et~al. (2006); 
and at $545\,$GHz and $857\,$GHz from the Herschel-ATLAS Science Demonstration 
Phase field (Clements et~al. 2010).  
The counts are multiplied by flux to the power $2.5$ to allow the 
contrast with the Euclidian counts to be better appreciated. 
} 
\label{DNDS}
\end{figure*}

\section{Results}\label{Results}

In this section we present results obtained using both the MCGAL and 
MILLGAL catalogues. We also show analytic predictions for the clustering 
of SFGs, and compare these to the more accurate predictions obtained from the 
MILLGAL catalogue, to highlight the shortcomings of the analytical 
calculations. We start by showing the predicted number counts of galaxies at 
the Planck frequencies to show how well the model reproduces the 
observed sub-millimetre counts (Sec.~\ref{counts}). Next we look at the 
contribution to the luminosity density from galaxies at different 
redshifts (Sec.~\ref{rhoL}). In Section~\ref{clustering} we show the predictions 
for the effective bias and contrast the analytical and direct estimates of the 
spatial correlation function of intensity. Finally, in Section~\ref{cl} 
we present the main predictions of the paper for the angular power 
spectrum of fluctuations in the CBL. 

\subsection{The number counts of galaxies at sub-millimetre wavelengths}
\label{counts}

We first show the model predictions for the number counts of galaxies 
in the Planck HFI bands for which we later present clustering predictions. 
Fig.~\ref{DNDS} shows the predicted differential galaxy counts 
and compares these with recent observational estimates. 
The top-right panel of Fig.~\ref{DNDS} is an update of the comparison 
shown by Baugh et~al. (2005) who considered the $850\,\mu{\rm m}$ number counts. 
Baugh et~al. showed that the MCGAL model reproduces the number counts and  
redshift distribution of $850\,\mu$m selected 
samples (see also Almeida et~al. 2011). The recent measurements of the number counts 
at $250\,\mu$m, $350\, \mu$m and $500\, \mu$m using the Science Demonstration Phase 
data from the Herschel Space Telescope by Clements et~al. (2011) and 
Oliver et~al. (2011) suggest that the model predicts too many sources 
at bright fluxes (see predictions in Lacey et~al. 2010). This is apparent from 
the comparison between model and observations in the other panels of Fig.~\ref{DNDS}. 

The predictions for fluctuations in the CBL are sensitive to the flux-weighted abundance of galaxies. For example, the expression for the 
shot noise contribution to the angular power spectrum (Eq.~12) depends 
on the square of the flux. This is similar to the weighting of $S^{2.5}$ applied 
to the differential number counts in Fig.~\ref{DNDS} (which is standard practice 
in the literature to expand the dynamic range of the counts). The dominant contribution 
to the shot noise is from fainter fluxes. The model predictions agree best with the 
observed counts at faint fluxes. The model overpredicts the counts at bright 
fluxes at $350\,\mu$m ($857\,$GHz) and $550\,\mu$m ($545\,$GHz) and underpredicts  the counts at bright fluxes at $1380\,\mu$m ($217\,$GHz), which 
is due to the neglect of radio galaxies in the model.

\subsection{The luminosity density of galaxies in the Planck bands}
\label{rhoL}

\begin{figure}
\includegraphics[width=8.6cm]{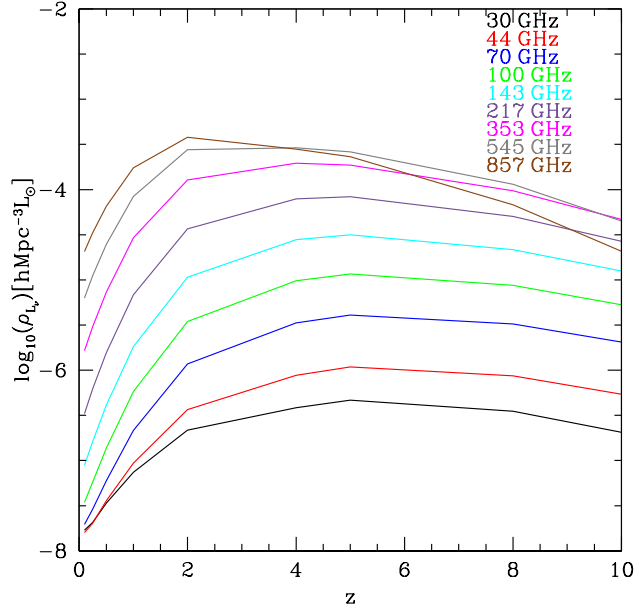}
\caption{
The luminosity density in the Planck wavebands as a function of redshift, 
predicted using the MCGAL catalogue. 
} 
\label{RHOLGRASIL}
\end{figure}

The luminosity density ($\rho_{L}$; Eq.~10) in the Planck frequency bands is plotted 
in Fig.~\ref{RHOLGRASIL} as a function of redshift, computed using all of the 
galaxies in the MCGAL catalogue. 
At all nine Planck frequencies, 
the luminosity density increases from the present day up to $z \approx 2$--$5$ 
and then stays approximately constant or declines gently to $z=10$. 
At a given redshift, the amplitude of the luminosity density increases 
with observer frame frequency because $L$ in the rest frame 
increases due to the shape of the SED (until the rest frame frequency moves 
past the peak in the dust emission spectrum). 
For a given observer frame frequency, the luminosity density is a combination  
of the SED sampled in the rest frame and the abundance of galaxies 
emitting at a given flux. The overall shape of the luminosity density 
as a function of redshift therefore tracks the star formation rate density 
in the universe, with a much gentler decline to high redshift, due to the 
increase in the rest frame $\nu L$ (due to the negative $k$-correction) 
offsetting the overall drop in star formation density.
 
The fraction of the overall luminosity density that is contributed by 
high redshift sources drops with increasing frequency, in agreement with 
previous interpretations of fluctuations in the CBL using empirical 
models (e.g. Hall et~al. 2010).  

\begin{figure}
\includegraphics[width=8.6cm]{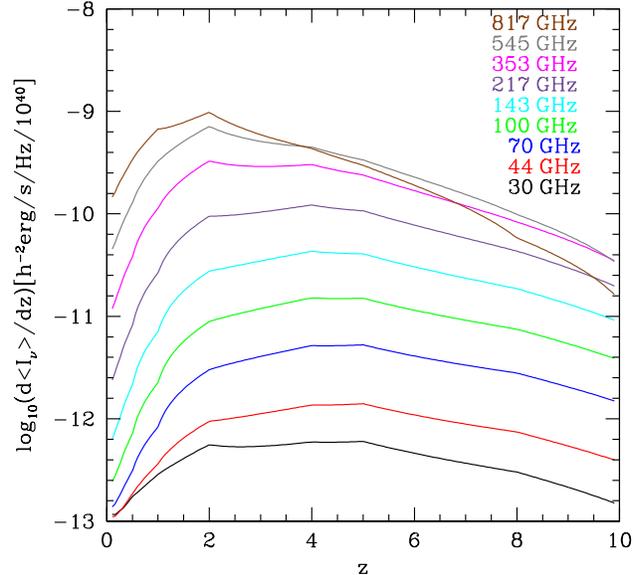}
\caption{
The contribution to the mean intensity per unit redshift interval 
in the Planck wavebands in the MCGAL catalogue (see Eq.~\ref{FZEQ}). 
} 
\label{FZ}
\end{figure}

\begin{figure}
\includegraphics[width=8.6cm]{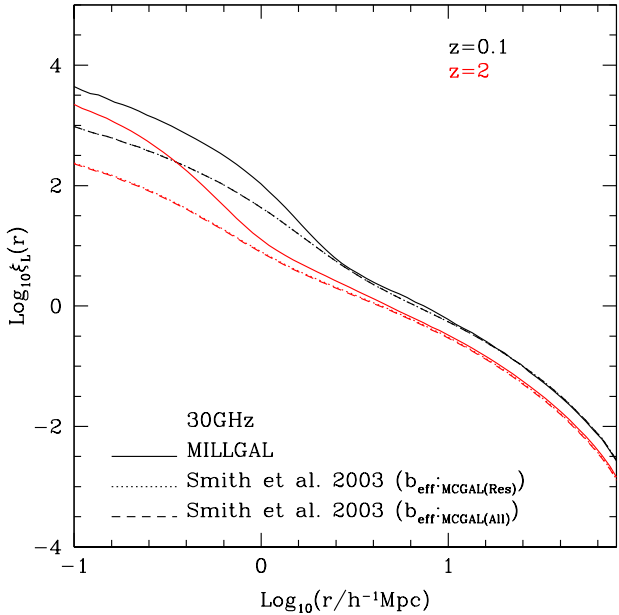}
\caption{
The two point luminosity correlation function at $z=0.1$ (black lines) and $z=2$ 
(red lines) for galaxies fainter than the Planck flux limit at $30$\,GHz. 
The direct estimates from the MILLGAL catalogue are shown by solid lines. 
The analytic calculations derived from the MCGAL sample are shown by dotted lines 
when using all galaxies and dashed lines when only using those galaxies which 
reside in haloes above the resolution limit of the Millennium. Note that 
the dotted and dashed lines coincide with one another at both redshifts, 
indicating that the finite resolution of the Millennium simulation has 
little impact on the predicted clustering of luminosity.
} 
\label{3DCFPlanck}
\end{figure}

\begin{figure}
\includegraphics[width=8.6cm]{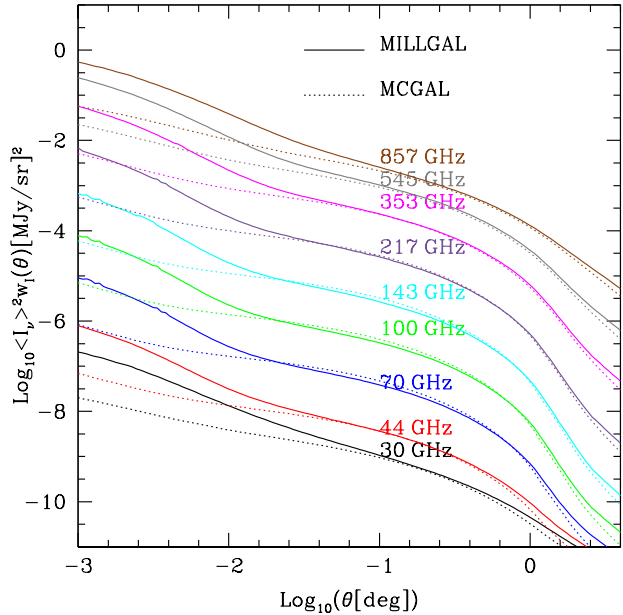}
\caption{
The product of the angular correlation function of intensity fluctuations and the 
square of the mean intensity for undetected galaxies in the nine Planck wavebands. 
Dotted lines are for galaxies in MCGAL catalogue 
using the analytical 
calculations of the bias factor and correlation function. 
Solid lines are for galaxies in MILLGAL catalogue, for which the clustering of 
luminosity is calculated using the Millennium N-body simulation.}  
\label{F2WFGN}
\end{figure}

The intensity fluctuation power spectrum depends on the mean 
intensity, which is an integral over the luminosity 
density as given by Eq.~\ref{FZEQ}.  
From this equation, the contribution to the mean intensity  
per unit redshift interval is given by: 
\begin{equation}
\frac{ {\rm d} \langle I \rangle}{ {\rm d} z } 
= \frac{1}{4 \pi} \frac{ \rho_{L} }{ (1 + z) } \frac{ {\rm d} x }{ {\rm d} z}.
\end{equation}
By plotting this quantity, we can see which redshifts contribute most to 
the mean intensity. 
Fig.~\ref{FZ} shows the contribution to the mean intensity with redshift 
in the Planck wavebands. 
A comparison between this plot and Fig.~\ref{RHOLGRASIL} 
shows that the mean intensity is dominated by lower redshifts than the 
luminosity density. 

\subsection{The clustering of SFGs in the Planck bands}
\label{clustering} 

We now turn our attention to the computation of the flux correlation function. 
The use of the Millennium N-body simulation allows us to make accurate 
predictions for the clustering of galaxies on small and intermediate scales. 
This is illustrated in Fig.~\ref{3DCFPlanck} which contrasts the direct 
estimate of the luminosity correlation function estimated from the 
MILLGAL catalogue with the analytic calculation based on the MCGAL catalogue. Recall that the pairwise 
galaxy counts are weighted by luminosity here. At small pair separations, 
$ r < 1 h^{-1}$Mpc, the N-body estimates are up to an order of 
magnitude larger than the analytic ones. Even though the analytic calculation 
takes into account the nonlinear form of the matter correlation function, the 
galaxy correlation function can be significantly different on these scales, 
since the analytic calculation ignores the 1-halo term (Benson et~al. 2000). 
The differences between the two estimates persist to intermediate separations of 
a few megaparsecs, in the transition region from the one-halo to two-halo 
contributions to the correlation function. 

In Fig.~\ref{F2WFGN} we take a step closer to the angular power spectrum of 
intensity fluctuations by plotting the angular flux correlation function 
multiplied by the square of the mean intensity fluctuations. 
We also compare the direct estimate of the clustering from the N-body simulation 
(solid lines) with the analytic one (dotted lines), 
for galaxies fainter than the Planck detection limits. 
The amplitude and shape of the two estimates of the correlation function 
are nearly the same at large angular separations. On small angular scales, e.g. 
$10^{-3}$ degrees, the two estimates differ by up to an order of magnitude, 
due to the more accurate treatment of the 1-halo term and the transition 
between the 1- and 2-halo regimes in the N-body calculation 
(as seen in Fig.~\ref{3DCFPlanck}). 

\begin{figure*}
\includegraphics[width=15cm]{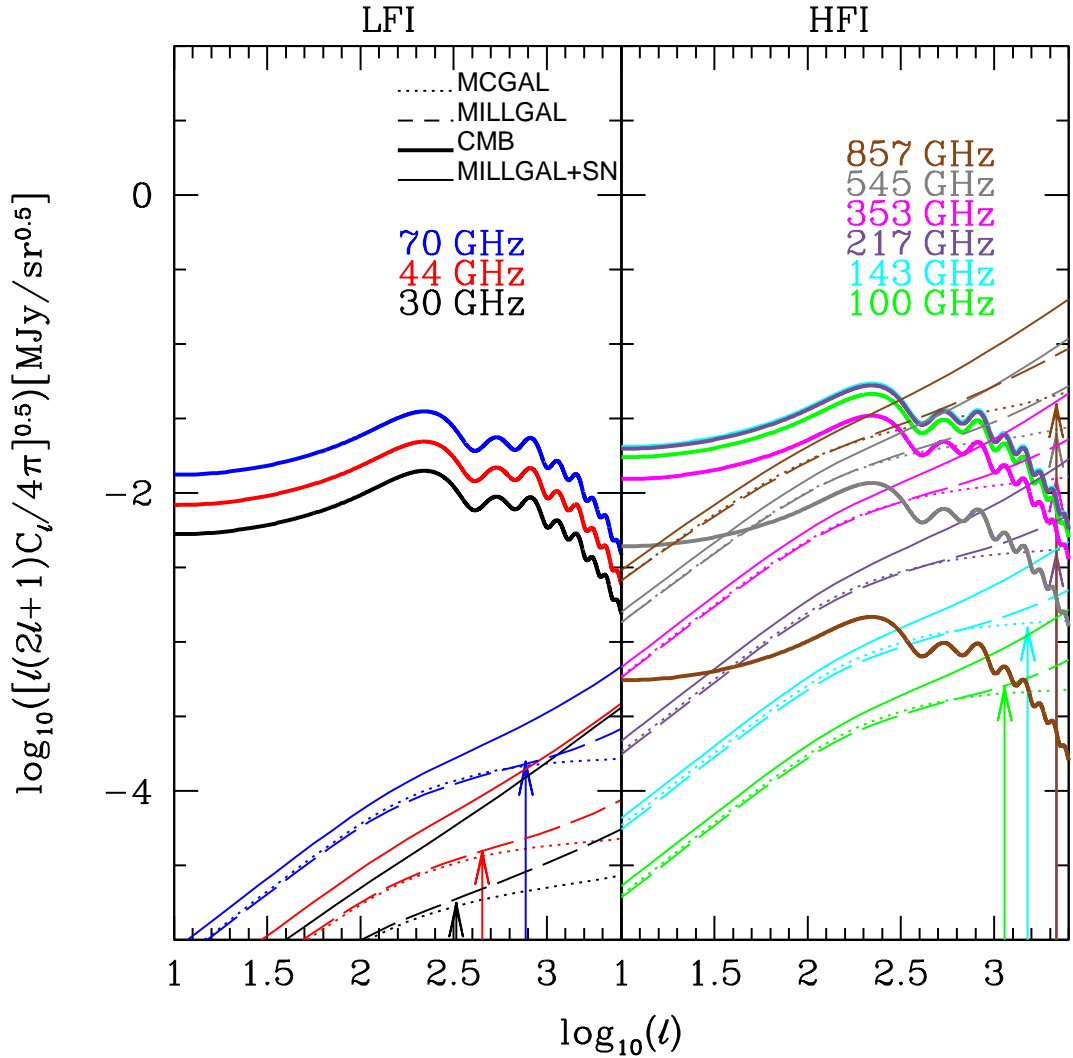}
\caption{
The angular power spectrum of the intensity fluctuations of undetected 
galaxies in the nine Planck wavebands. Different colours show the frequency 
bands as listed in the key. The left panel shows the LFI frequencies and 
the right panel shows the HFI wavebands. 
The dotted lines show the analytic estimates of the intrinsic clustering, 
which should be compared with the estimates made using the Millennium simulation, 
which are shown by the long-dashed lines. The thin solid lines show the full 
prediction for the clustering, combining the intrinsic clustering with the 
shot noise derived from the number counts of unresolved sources. 
The thick lines show the primordial CMB power spectra. The vertical arrows 
indicate the angular resolution of Planck listed in Table~1 and are colour coded by 
frequency.}
\label{ANGPOWER}
\end{figure*}

\begin{figure*}
\includegraphics[width=8.6cm]{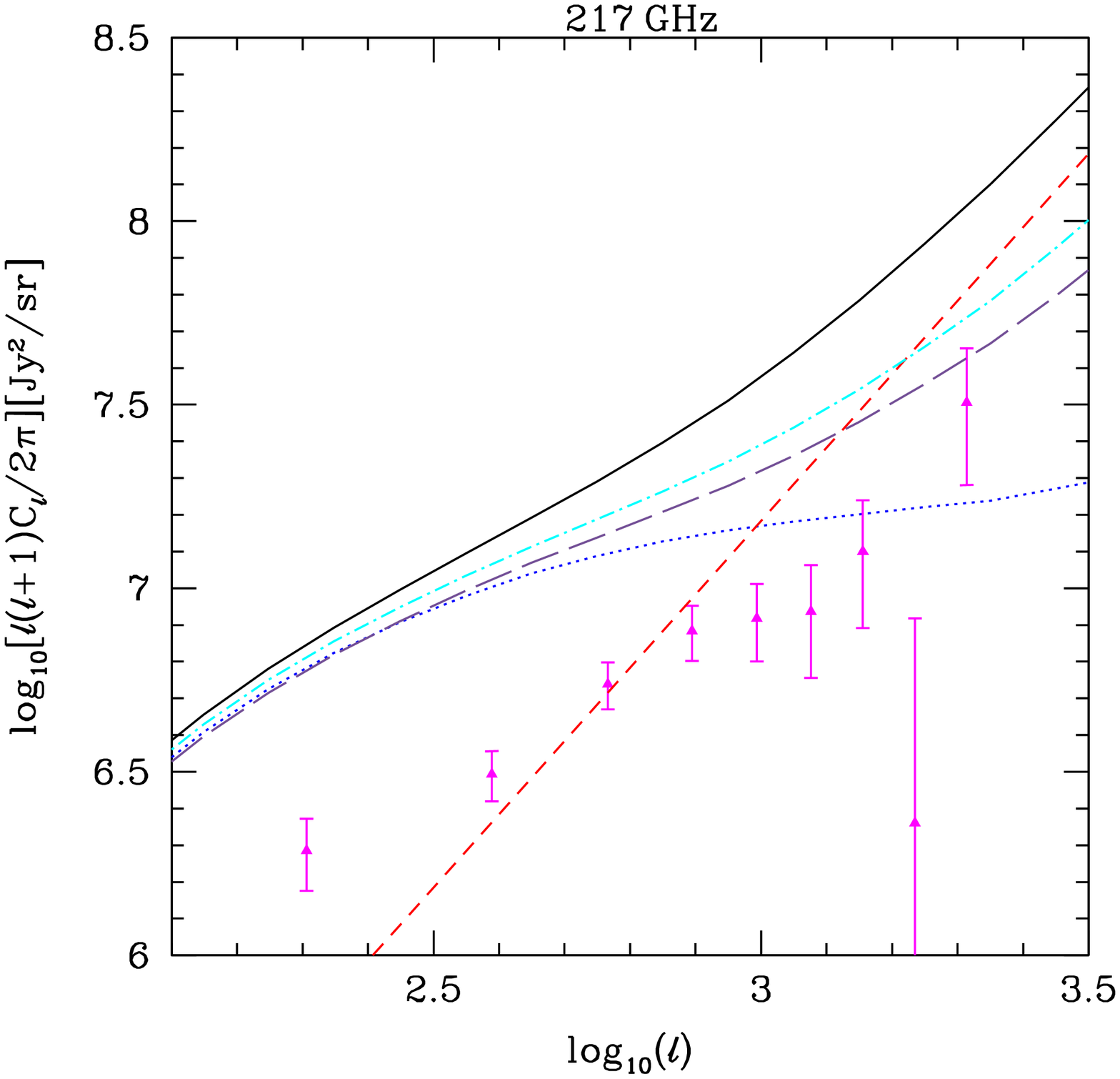}
\includegraphics[width=8.6cm]{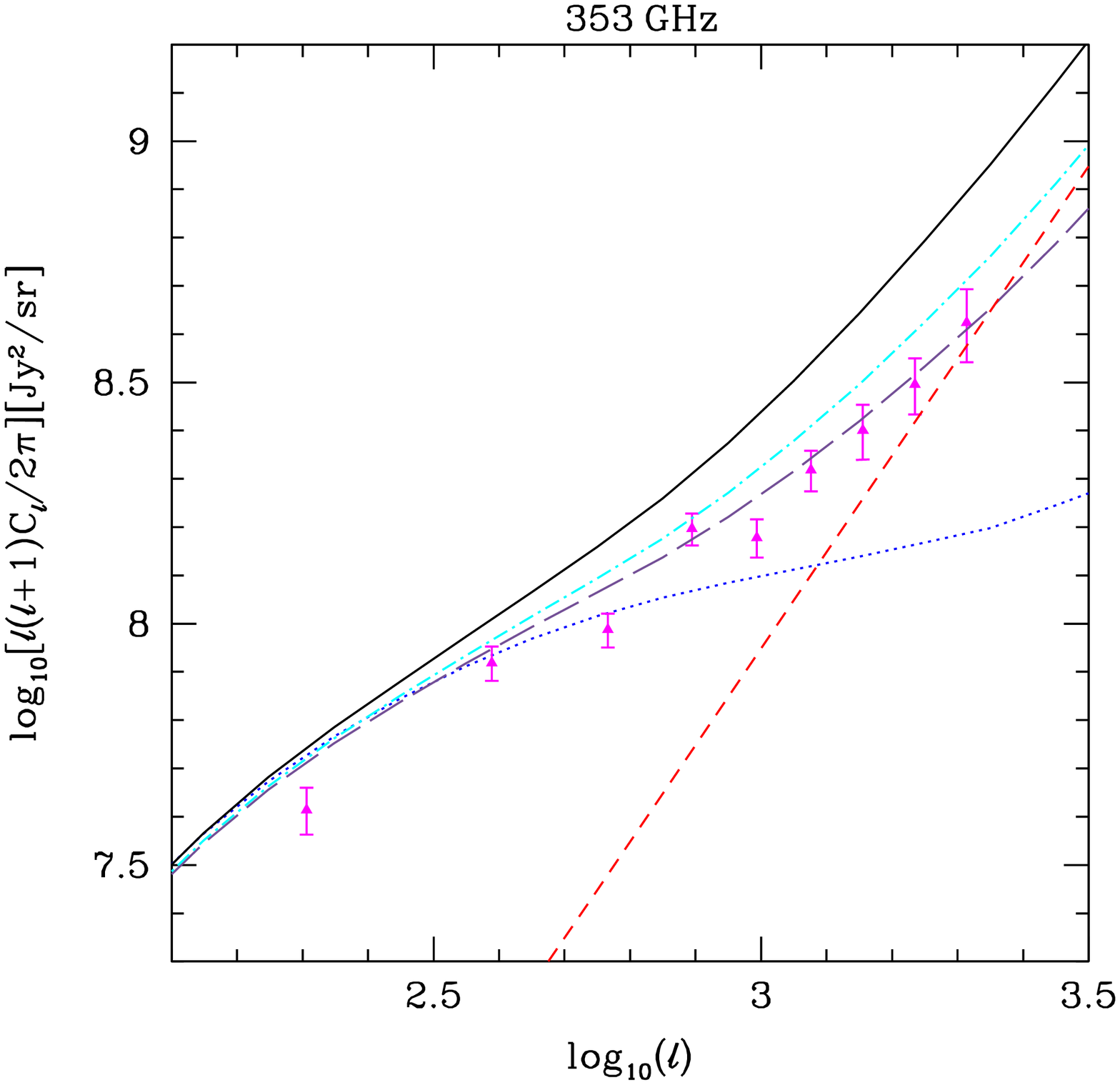}
\includegraphics[width=8.6cm]{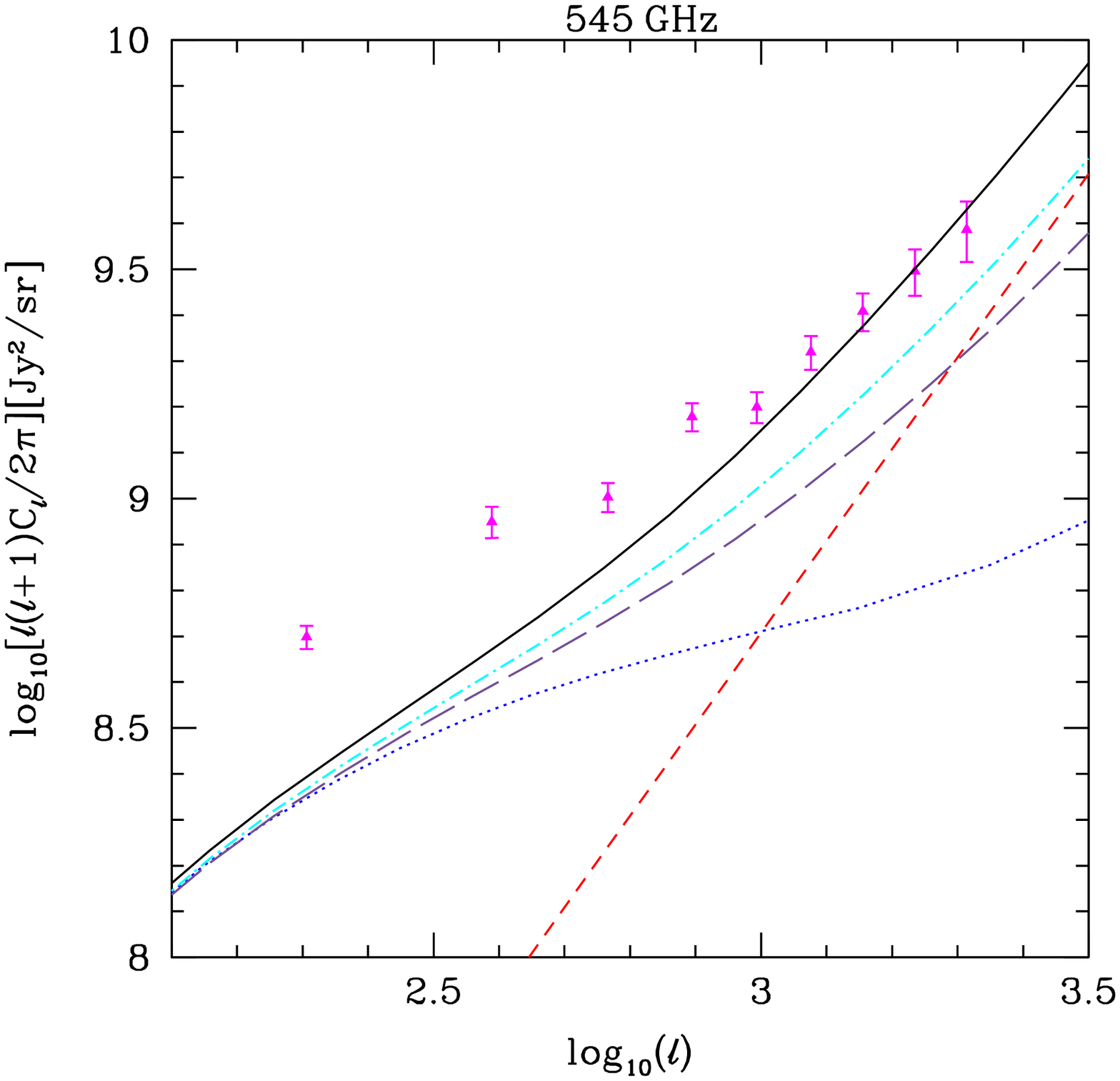}
\includegraphics[width=8.6cm]{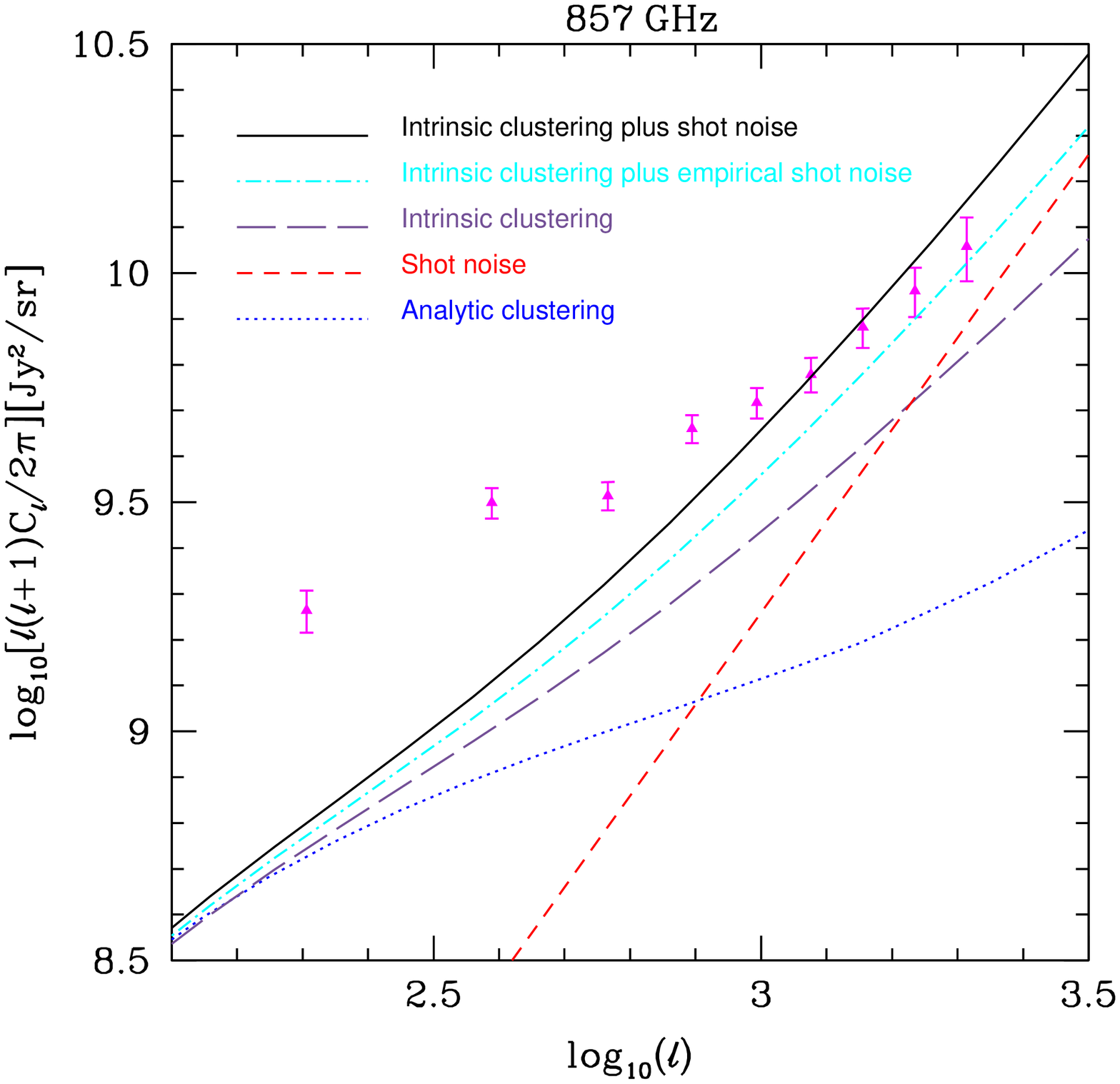}
\caption{
The angular power spectrum of intensity fluctuations in the extragalactic 
cosmic infrared and millimetre background. Each panel corresponds to a different HFI 
frequency as indicated. The points show the extragalactic background fluctuation 
estimated from Planck measurements (The Planck Collaboration 2011b), after 
removing the cosmological signal and the Galactic foreground. The error bars 
show the statistical and systematic errors. The lines show our theoretical predictions. The 
short-dashed line shows the shot noise from the counts predicted by the 
MILLGAL model. The long dashed line shows the clustering estimated 
from the MILLGAL catalogue, without any shot noise, which we refer to as the 
intrinsic clustering. The analytic estimate of the intrinsic clustering is shown 
by the dotted line. The solid line shows our full prediction for the intensity 
fluctuations, combining the intrinsic clustering (long dashed line) and the 
shot noise (short dashed line). Another version of this prediction, derived 
by combining the intrinsic clustering with the shot noise estimated using the 
model of Bethermin et~al. (2011) is shown by the dot-dashed line. 
}
\label{ANGPOWERDATA}
\end{figure*}

\subsection{The angular power spectrum of fluctuations in the CBL from SFGs}
\label{cl}

The main results of the paper are shown in Figs.~\ref{ANGPOWER} and \ref{ANGPOWERDATA}.
Fig.~\ref{ANGPOWER} shows the angular power spectrum of the intensity fluctuations of 
undetected galaxies in the Planck wavebands. Different components of the 
model predictions are shown in this plot. The long-dashed line shows the intrinsic 
clustering predicted using the N-body simulation, without any contribution 
from shot noise.  This is to be contrasted with the dotted lines, which show 
the analytic clustering estimate. At small angular scales, $l > 3000$, the N-body 
estimate exceeds the analytic one. 
Fig.~\ref{ANGPOWER} also lets us compare the amplitude of 
the fluctuations from extragalactic sources to the primordial CMB signal. In the LFI  
channels, the primordial signal dominates on all angular scales. At the HFI frequencies, 
the extragalactic and primordial signals become comparable above a particular value of 
$l$. 
The angular power spectrum from undetected faint extragalactic sources 
exceeds the primordial power spectrum from $l \sim 1000$ at $353$~GHz, 
$l \sim 100$ at $545~$GHz and for all angular scales at $857~$GHz. 

Early measurements by Planck have been used to estimate the fluctuations in 
the CBL from unresolved extragalactic sources (The Planck Collaboration 2011b). 
As we have seen in Fig.~\ref{ANGPOWER}, the primordial signal is orders of 
magnitude larger than the extragalactic signal at the lowest frequencies measured 
by Planck. Maps at these frequencies can be used to estimate the primordial signal 
expected in the higher frequency bands. These authors remove the Galactic foreground 
using observations of emission from neutral hydrogen to clean thermal dust emission. 
The estimated fluctuations in the cosmic infrared background due to the unresolved 
extragalactic sources are shown by the data points in Fig.~\ref{ANGPOWERDATA}. The 
error bars show the statistical and systematic errors (the area used covers 140 
square degrees in six different fields). The systematic error, e.g. due to 
uncertainties in the beam, exceeds the statistical error at high $l$ in the 
higher frequency channels (see the Planck Collaboration 2011b for further details). 

The theoretical predictions for the extragalactic intensity fluctuations in the CBL are considered more closely in Fig.~\ref{ANGPOWERDATA}. 
The shot noise computed from the MILLGAL model using 
Eq.~\ref{shot} is shown by the short dashed line. The shot noise makes a fixed 
contribution to $C_{l}$ but has a scale dependence in Fig.~\ref{ANGPOWERDATA} since 
here we plot $ l ( l + 1) C_{l}$. The contribution to $C_{l}$ from the 
intrinsic clustering of the unresolved galaxies, as estimated using the 
N-body based MILLGAL catalogue, is shown by the long dashed line. The corresponding 
analytical estimate is shown by the dotted line. At the smallest scales plotted, the 
N-body estimate exceeds the less accurate analytical one by a factor of three or more. 
The overall prediction for the intensity fluctuation power spectrum, combining the 
intrinsic clustering with the shot noise, is shown by the solid line. 
The shot noise makes an important contribution to the power spectrum at 
high $l$. We have commented already that our model overpredicts the number 
counts of galaxies in the Herschel bands at bright fluxes. This will result in turn in an 
overprediction of the shot noise in these bands. To illustrate how this can 
affect our predictions, we also show a version of the power spectrum 
in which we replace the shot noise calculated using our model with that from 
the empirical number count model of Bethermin et~al. (2011). This alternative 
prediction is shown by the dot-dashed line in Fig.~\ref{ANGPOWERDATA}. 
We note that these predictions are not meant to supersede those shown by 
the solid line, as now the number of sources is decoupled 
from the calculation of the intrinsic clustering. 

Fig.~\ref{ANGPOWERDATA} shows that the model predictions are generally within a 
factor of three or better of the observational estimate of the extragalactic 
background fluctuations. It is notable that in {\bf some} of the HFI bands, 
the predicted 
shape of the power spectrum is similar to the observational estimate. 
The agreement between the model and observations is best at small angular scales 
in the two highest frequency channels. There is a mismatch in amplitude on 
larger scales (smaller $ l$) in these higher frequency channels.

\subsection{How can the model predictions be improved?}

The goal of this paper is to present a new framework for 
predicting the contribution of star-forming galaxies to 
intensity fluctuations in the cosmic background light. 
We have used a previously published model to illustrate 
the technique. Unlike other studies in the literature, we 
have not varied any model parameters in order to improve 
the agreement of the model predictions with the signal 
inferred from observations. 

Fig.~\ref{ANGPOWERDATA} shows the model underpredicts the 
observed clustering on large angular scales in the two 
highest frequency HFI channels, whilst giving a reasonable 
match to the power spectrum on small scales. To improve the 
model predictions at $545$\, GHz and $857$\,GHz, we would need 
to increase the effective bias of the unresolved sources, whilst 
reducing their number slightly. This requires an increase in the 
typical effective host halo mass. Such a shift could be achieved 
by reducing the efficiency of star formation in galaxies hosted 
by low mass haloes. 

On the other hand in the two lowest frequency channels shown in 
Fig.~\ref{ANGPOWERDATA}, $217$\,GHz and $353$\,GHz, the model overpredicts 
the clustering, so there needs to be a reduction in the effective host halo 
mass in these cases. We note that the model works best overall at 
$353$\,GHz, the frequency at which the model was originally tuned to 
match the observed galaxy number counts and redshift distribution.

\begin{figure}
\includegraphics[width=8.6cm]{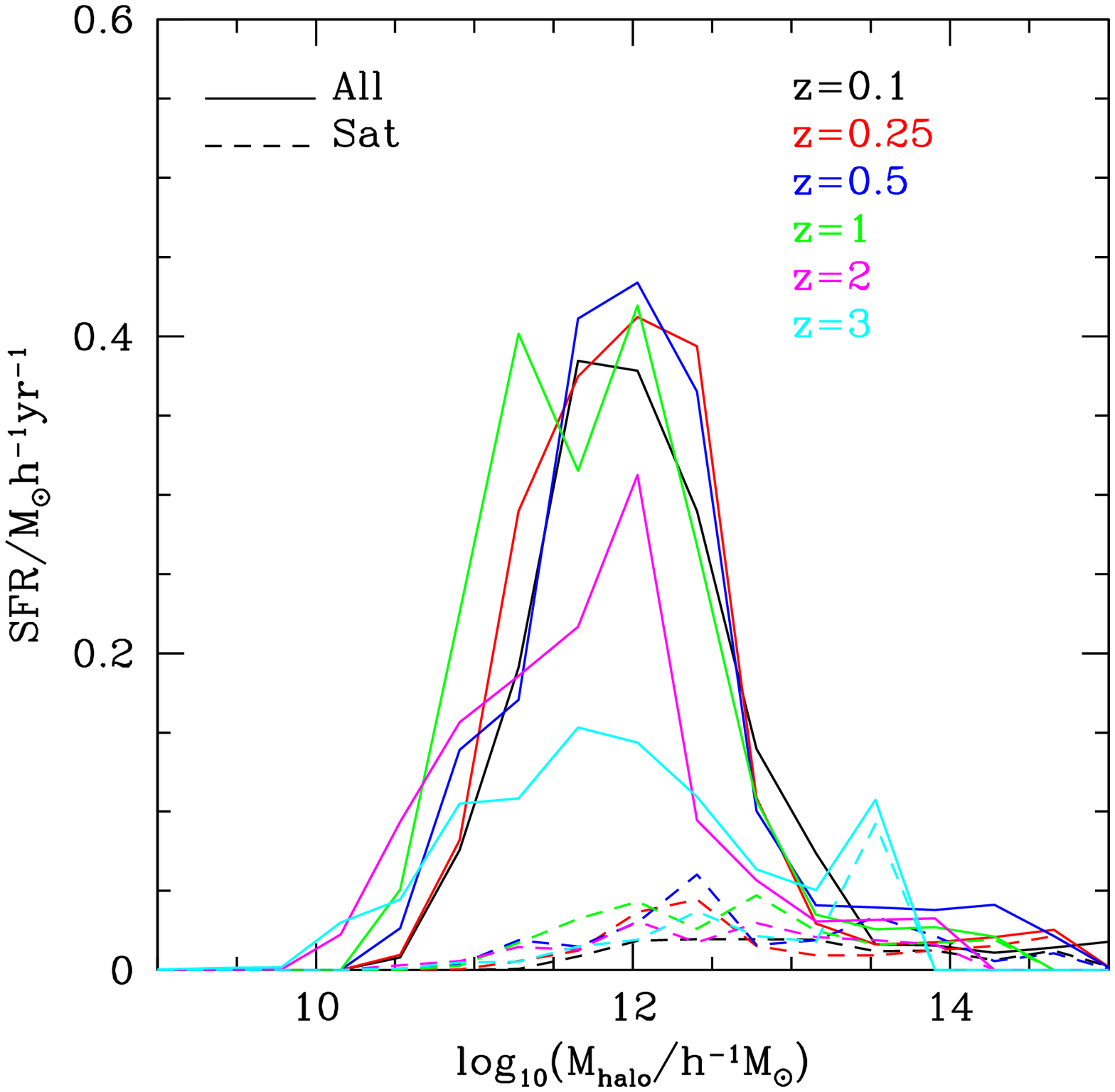}
\includegraphics[width=8.6cm]{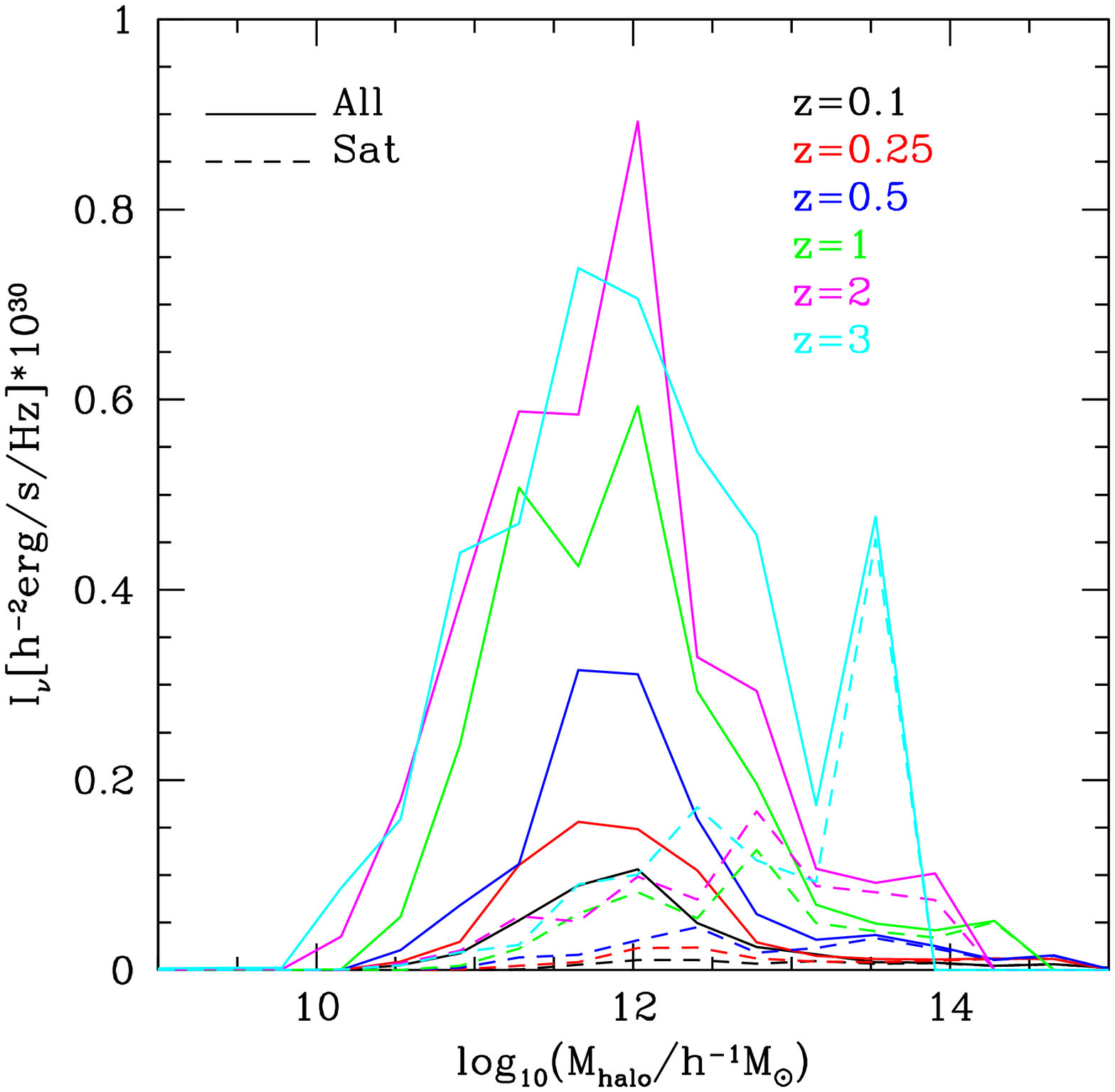}
\caption{
Top panel: the total star formation 
rate summing over galaxies within common 
dark matter haloes as a function of host halo 
mass. Different redshifts are shown by different 
coloured lines, as indicated by the key. The solid 
lines show the predictions for all galaxies and the 
dashed lines for satellite galaxies only. 
Bottom panel: the total luminosity per halo in the 
observer frame at $353$\,GHz 
as a function of host halo mass. Lines styles 
have the same meaning as in the top panel.}
\label{REFEREE}
\end{figure}

Fig.~\ref{REFEREE} gives some insight into which haloes 
dominate the clustering signal. The top panel of Fig.~\ref{REFEREE} 
shows the total star formation rate summing over galaxies in 
the same dark matter halo, plotted as 
a function of host halo mass. The y-axis is on a linear scale 
so that we can gain an impression of which haloes make the most important 
contribution to the overall star formation rate density (we would also 
need to take into account the halo mass function to connect this plot 
to the luminosity density).  
The distribution is dominated by central galaxies 
hosted by dark matter haloes with masses in the range 
$10^{11}$-$10^{12 }\,h^{-1}\,M_{\odot}$. In this paper 
we study correlations in intensity, so an important 
quantity to consider is the total luminosity per halo, summing over 
galaxies within the same dark matter halo, which is plotted as a 
function of host halo mass in the bottom panel of Fig.~\ref{REFEREE}. 
The far-infrared luminosity depends on a galaxy's star formation rate, along with its 
dust content and the level of extinction. The distribution of luminosity per halo  
has a similar form to that of the star formation rate, with a 
more pronounced tail to higher halo masses. This plot shows that 
the mass resolution of the Millennium N-body simulation is sufficient 
for modelling the intensity fluctuations of unresolved galaxies. 
To change the clustering predictions of the model, we need 
to move the location of the peak in the distribution plotted 
in Fig.~\ref{REFEREE}, which means finding a way to put central 
galaxies of a given luminosity into different mass haloes, depending 
on the sense of the change required in the two-halo clustering 
term. The tail of satellite galaxies apparent in higher mass haloes 
influences the one-halo clustering term (i.e. pairs of galaxies 
within the same halo).

We have carried out a preliminary exploration of parameter space, changing 
one aspect of the model at a time, without trying to retain the previous successes 
of the model. We altered the strength of feedback from supernovae and experimented 
with deleting satellite galaxies assigned to subhaloes which can no longer 
be resolved in the Millennium Simulation. These changes produce a similar shift 
in the model predictions at each frequency and so do not produce the differential 
changes that we need to improve the match to the observations. A full, multi-dimensional 
parameter search to find a model with an improved match to the intensity 
fluctuations, which also reproduces the previous successful matches to observations, 
is beyond the scope of the current paper.

\section{Conclusions}
\label{Summary}

Fluctuations in the intensity of the microwave background radiation arise from a number of 
sources: ripples in the density of matter in the primordial Universe, emission from the 
interstellar medium in our galaxy, and extragalactic sources, such as star-forming 
galaxies (SFGs), radio galaxies or the hot plasma in galaxy clusters. Identifiable sources 
can be removed from intensity fluctuation maps. Sources below the confusion  limits of the 
measurements cannot be explicitly excised. Their contribution can be removed statistically 
in analyses of the primordial signal or can be isolated to study the history of star formation 
in the Universe and its connection to structure formation in the dark matter. 

Here we have introduced a hybrid model which combines a physical model of galaxy formation 
with a N-body simulation of the clustering of dark matter to predict the contribution 
of SFGs to the intensity fluctuations in the CBL. This is the first time that it has been 
possible to compute the abundance and clustering of SFGs together in a physical model. 
The model predicts the radio emission from star-forming galaxies and the emission from 
dust heated by stars, in a self consistent manner, with the dust grains in thermal 
equilibrium. The amount of heating depends on the star formation and chemical enrichment 
history of the galaxies, and on their dust mass and linear size; all of these properties 
are predicted by the model. Our calculation tells us how many galaxies above a stated flux 
limit should reside within dark matter haloes of a given mass. This information, in combination 
with an N-body simulation of the clustering of dark matter, allows a direct calculation of the 
clustering of SFGs. In previous calculations based on dark matter haloes, the halo occupation 
distribution was adjusted by hand, with no connection between the 
properties of the host halo and the spectral energy distribution of the galaxy. 

We have illustrated this new framework using a published model of galaxy formation 
(Baugh et~al. 2005). This model matches the observed galaxy number counts at 353 GHz (850 \,$\mu$m), 
along with other observations of the galaxy population at low and high redshift. We have taken 
the predictions of this model without any adjustments. Hence, the example calculation we present 
is in effect parameter-free, since we make no further adjustment to the values of the parameters 
which specify the galaxy formation model. This is an important distinction of our work from 
halo occupation distribution modelling of the CBL fluctuations, in which case the model parameters 
are adjusted to improve the match to the measured clustering. In view of this, it is 
impressive that our predictions only disagree with the fluctuations in the CBL inferred 
from early Planck data by at worst a factor of three. It is also 
important to bear in mind that the observational estimate of the CBL fluctuations is heavily 
processed, and significant contributions from other sources have to be removed to isolate the 
extragalactic signal. Empirical calculations in the literature have a more limited scope than 
our model. Such calculations do contain parameters which are tuned to improve directly the 
agreement between the predicted and observed clustering. However, {\bf in general} these models do not connect 
the emission from galaxies to their host dark matter haloes (for an 
exception see e.g. Shang et~al. 2012). 

Our model agrees best the inferred CBL fluctuations at high frequencies 
on small scales (high $l$), for which the shot noise from discrete sources 
and the clustering of galaxies within common dark matter haloes dominate, and 
where the isolation of the extragalactic signal is most secure.  
By using an N-body simulation, we are able to make accurate predictions 
for the one-halo as well as two halo contribution to the intensity fluctuations. These 
predictions are significantly different from simple analytical calculations 
on small angular scales. In general the model does less well on larger scales, 
around $ l \sim 100$. This implies that the example model predicts the wrong effective 
bias or two-halo clustering term. At $217$GHz, the predicted bias is too high by a 
factor of $1.6$.  At $857$GHz, the effective bias predicted is too small by a similar 
factor ($1.7$). As the intrinsic clustering dominates over the shot noise on these scales,  
this suggests that some redistribution of galaxies between dark matter 
haloes is required in the model. This is more difficult to realise than it may sound, as these 
adjustments would have to be made without changing the small-scale clustering by much. Of course, 
if the number of sources is also changed, then this will alter the shot noise, which mainly 
affects the amplitude of the small scale clustering.  Another possible 
explanation for the discrepancy between the model predictions and the 
Planck results is that the emissivity of the dust assumed in the model 
may be incorrect. Baugh et~al. modified the dust emissivity in bursts 
from the standard value of $\epsilon \propto v^{-2}$ used in quiescent 
star formation to $\epsilon \propto v^{-1.5}$. This boosts the emission 
at longer wavelengths, and may in part be responsible for the excess counts 
at low frequencies. 

As new observations of SFGs become available through, for example, the 
Herschel Space Observatory, new constraints will be placed on the 
galaxy formation model which underpins our method (Lacey et~al. 2008; 2010). Along with 
improved treatments of key model ingredients, such as star formation (Lagos et~al. 2010), 
this will allow us to devise new models which better match the new observations of SFGs. 
The clustering of intensity fluctuations in the CBL will offer an important additional 
observational constraint that such galaxy formation models will be able to exploit.

\section*{Acknowledgments}
We acknowledge a helpful report from the referee. 
HSK acknowledges support from the Korean Government's Overseas Scholarship. 
CSF acknowledges a Royal Society Wolfson Research Merit Award. 
SMC acknowledges the support of a Leverhulme Research Fellowship. 
This work was supported in part by a rolling grant from the Science and 
Technology Facilities Council.

\section*{Appendix}
\label{APP}
The halo mass resolution of the {\tt MCGAL} catalogue is essentially arbitrary, as 
the full memory of the computer can be devoted to a single dark matter halo 
merger history. Furthermore, the mass resolution can be adjusted with redshift, 
to ensure that a representative sample of halo masses is modelled at each epoch. 
The mass resolution of the {\tt MILLGAL} sample is set by the Millennium 
N-body simulation and is fixed at all redshifts. In this Appendix we compare results 
from the two calculations to demonstrate the impact that the finite resolution of 
the {\tt MILLGAL} sample has on our model predictions. We conclude that the predictions 
for the intensity fluctuations are insensitive to the resolution of the Millennium 
Simulation. 

\begin{figure}
\includegraphics[width=8.6cm]{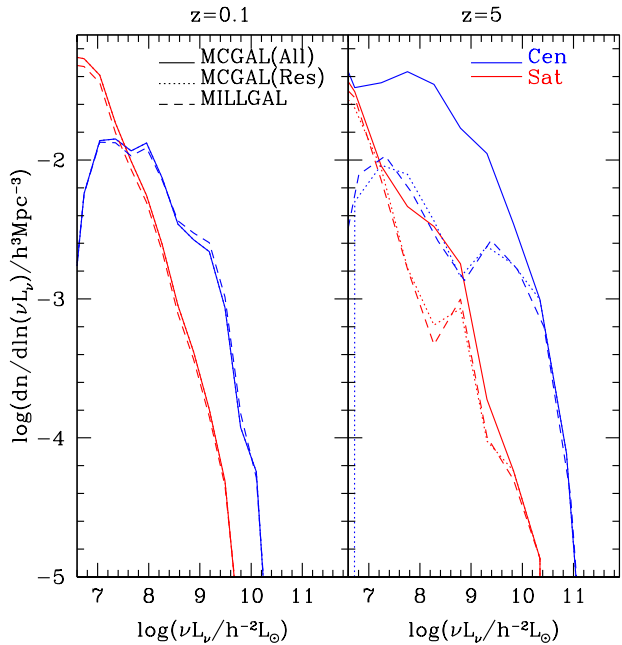}
\caption{
The luminosity function (LF) at 857~GHz (350\,$\mu$m) at $z=0.1$ (left) and $z=5$ (right). 
Solid lines shows the full MCGAL catalogue and dotted lines show the result from this 
catalogue when restricted to haloes with masses above the resolution limit of the Millennium 
simulation ($M_{\rm halo} = 1.72 \times 10^{10} h^{-1} {\rm M_{\odot}}$). 
Note that the dotted and solid lines are coincident in the left panel. 
The dashed line shows the predictions from the MILLGAL catalogue. 
The blue lines show the contribution to the LFs from central galaxies. 
The red lines show the LF of satellite galaxies.} 
\label{LFGN}
\end{figure}

We first consider the luminosity functions predicted in the two catalogues 
in Fig.~\ref{LFGN}. At low redshift, the MILLGAL and MCGAL catalogues are 
in very good agreement with one another, both for central and satellite galaxies 
(i.e. at $z=0.1$ in the left panel of Fig.~\ref{LFGN}). 
The luminosity function of the two catalogues 
differ at high redshift because the MILLGAL does not 
include galaxies hosted by haloes below the mass resolution 
of the Millennium simulation. However, the MILLGAL catalogue reproduces well the 
luminosity function of the MCGAL catalogue at higher luminosities, at which 
the galaxies tend to be hosted by more massive dark matter haloes which 
are resolved in Millennium simulation (see the right panel in Fig.~\ref{LFGN}). 

\begin{figure}
\includegraphics[width=8.6cm]{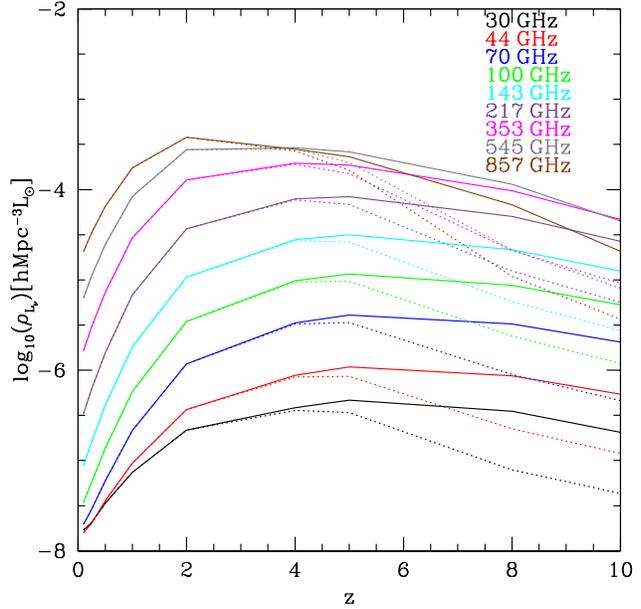}
\caption{
The luminosity density in the Planck wavebands as a function of redshift, 
predicted using the MCGAL catalogue. The solid lines show the predictions 
using all galaxies in the MCGAL catalogue and the dotted lines show the 
results using only those galaxies hosted by haloes which could be resolved 
in the Millennium Simulation.
} 
\label{RHOO}
\end{figure}

In Fig.~\ref{RHOO}, the luminosity density of galaxies hosted by dark matter haloes 
resolved by the Millennium simulation is lower than that of all 
the galaxies in the MCGAL catalogue for $z > 4$.  Galaxies in low mass dark 
matter haloes contribute significantly to the luminosity density at high $z$. 
 
\begin{figure}
\includegraphics[width=8.6cm]{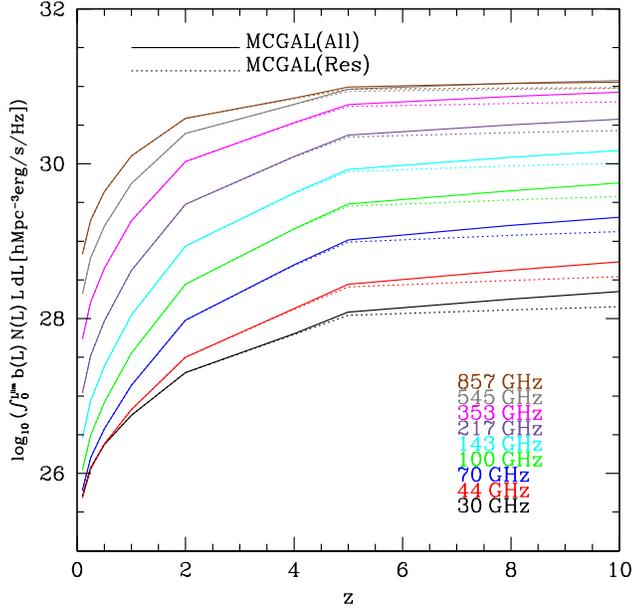}
\caption{
The integrated luminosity weighted bias (the numerator of the effective bias as defined 
in Eq.~11) as a function of redshift in the Planck wavebands calculated 
using the analytic halo bias from Sheth et~al. (2001). Only galaxies fainter 
than the Planck detection limits listed in Table~1 contribute. The solid curves 
show the predictions using all the galaxies in the MCGAL catalogue. The dotted 
curves show the results using only those galaxies hosted by haloes which could 
be resolved in the Millennium simulation. 
} 
\label{BeffR}
\end{figure}

In Fig.~\ref{BeffR}, we plot the integrated luminosity weighted bias. The result 
shows that there is little impact on this measure of the intensity fluctuations 
on applying the resolution limit of the N-body simulation to the MCGAL catalogue. 

\begin{figure}
\includegraphics[width=8.6cm]{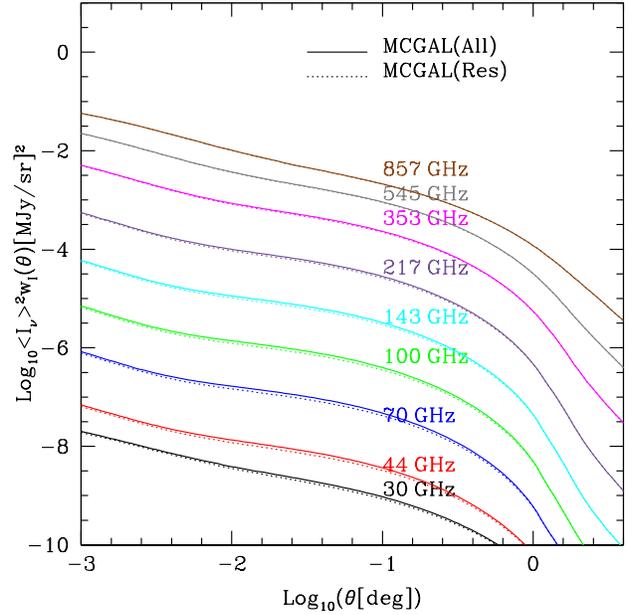}
\caption{
The product of the angular correlation function of intensity fluctuations and the 
square of the mean intensity for undetected galaxies in the nine Planck wavebands. 
Solid lines are for all galaxies in MCGAL catalogue and dotted lines are for galaxies 
in haloes which could be resolved in the Millennium simulation, using the analytical 
calculations of the bias factor and correlation function. 
These predictions are very similar, showing that halo mass resolution has little impact 
on our predictions. 
} 
\label{F2WFGNR}
\end{figure}

The similarity between the results shown in Fig.~\ref{BeffR} for different 
halo mass resolution limits and the fact that the mean intensity is 
dominated by low redshifts imply that our predictions for the power spectrum 
of intensity fluctuations should be insensitive to the resolution limit of the MCGAL 
catalogue. This is confirmed in Fig.~\ref{F2WFGNR} in which we plot the analytic 
estimate of the product of the square of the mean intensity and the angular 
correlation function of intensity using the MCGAL catalogue. The predictions 
for the full MCGAL catalogue are indistinguishable from those restricted 
to galaxies which could be resolved in the Millennium simulation.

\begin{thebibliography}{}

\bibitem{}
Almeida C., Baugh C.M., Lacey C.G., 2011, MNRAS, 417, 2057

\bibitem{}
Amblard A., Cooray A., 2007, ApJ, 670, 903


\bibitem{}
Angulo R.E., Baugh C.M., Lacey C.G., 2008, MNRAS, 387, 921

\bibitem{Barger}
Barger A. J., Cowie L. L., Sanders D. B., 1999, ApJ, 518, 5

\bibitem{Baugh}
Baugh C. M., Lacey C. G., Frenk C. S., Granato G. L., Silva L., Bressan A., Benson A. J., Cole S., 2005, MNRAS, 356, 1191

\bibitem{Baugh2006}
Baugh C. M., 2006, RPPh, 69, 3101

\bibitem{Benson1}
Benson A. J., Cole, S., Frenk C. S., Baugh C. M., Lacey C. G. 2000, MNRAS, 311, 739

\bibitem{Berlind1}
Berlind A. A., Weinberg D. H., 2002, ApJ, 575, 587

\bibitem{Bertoldi}
Bertoldi F.~et al., 2000, A\&A, 360, 92

\bibitem{}
Bethermin M., Dole, H., Beelen, A., Aussel, H., 2010, A\&A, 512, 78

\bibitem{}
Bethermin M., Dole, H., Lagache, G., Le Borgne D., Penin A., 2011, A\&A, 529, 4
 
\bibitem{Borys}
Borys C., Chapman S., Halpern M., Scott D., 2003, MNRAS, 344, 385

\bibitem{}
Bower R.G., Benson A.J., Malbon R., Helly J.C., Frenk C.S., 
Baugh C.M., Cole S., Lacey C.G., 2006, MNRAS, 370, 645

\bibitem{}
Bressan A., Silva L., Granato G. L., 2002, A\& A, 392, 377 

\bibitem{Chapman}
Chapman S. C., Scott D., Borys C., Fahlman G. G., 2002, MNRAS, 330, 92

\bibitem{}
Clements D.L., et al. 2010, A\&A, 518, L8

\bibitem{Cole}
Cole S., Lacey C. G., Baugh C. M., Frenk C. S., 2000, MNRAS, 319, 168

\bibitem{}
Condon J. J., ARA\&A, 30, 575. 


\bibitem{}
Coppin K., et~al. 2006, MNRAS, 372, 1621

\bibitem{Cowie}
Cowie L. L., Barger A. J., Kneib J.-P., 2002, AJ, 123, 2197

\bibitem{Dole2001}
Dole H.~et al., A\&A, 372, 364

\bibitem{Dole2004}
Dole H.~et al., 2004, ApJS, 154, 93


\bibitem{Dunkley}
Dunkley J., et al., 2011, ApJ, 739, 52

\bibitem{Elbaz}
Elbaz D., Cesarsky C. J., Chanial P., Aussel H., Franceschini A., Fadda D., Chary R. R., 2002, A\&A, 384, 848

\bibitem{eke}
Eke V., Cole S., Frenk C.S., Navarro J.F., 1996, MNRAS, 281, 703


\bibitem{Fernandez}
Fernandez-Conde N., Lagache G., Puget J.-L., Dole H., 2008, A\&A, 481, 885

\bibitem{Gao}
Gao L., Springel V., White S.D.M., 2005, MNRAS, 363, L66


\bibitem{Genzel}
Genzel R., Cesarsky C. J., 2000, ARA\&A, 38, 761

\bibitem{Gonzalez-Neuvo}
Gonzalez-Nuevo J., Toffolatti L., 2005, ApJ, 621, 1

\bibitem{Granato2000}
Granato G. L., Lacey C. G., Silva L., Bressan A., Baugh C. M., Cole S., Frenk C. S., 2000, ApJ, 542, 710


\bibitem{Granato2004}
Granato G. L., De Zotti G., Silva L., Bressan A., Danese L., 2004, ApJ, 600, 580


\bibitem{}
Grossan B, Smoot G.F., 2007, A\&A, 474, 731
 
\bibitem{Haiman}
Haiman Z., Knox L., 2000, ApJ, 530, 124

\bibitem{}
Hall N.R., et~al. 2010, ApJ, 718, 632. 

\bibitem{Holland}
Holland W. S., Greaves J. S., Zuckerman B., Webb R. A., McCarthy C., Coulson I. M., Walther D. M., Dent W. R. F.~et al., 1998, Nature, 392, 788

\bibitem{kennicutt1983}
Kennicutt R. C. Jr., 1983,  ApJ, 272, 54

\bibitem{kennicutt1998}
Kennicutt R. C. Jr., 1998, ARA\&A, 36, 189

\bibitem{Kim}
Kim H.-S, Baugh C. M., Cole S., Frenk C. S., Benson A. J., 2009, MNRAS, 400, 1527

\bibitem{}
Knox L., Cooray A., Eisenstein, D., Haiman Z., 2001, ApJ, 550, 7



\bibitem{Komatsu}
Komatsu E., et~al. 2011, ApJS, 192, 18.

\bibitem{kroupa}
Kroupa P., Aarseth S., Hurley J., 2001, MNRAS, 321, 699

\bibitem{lacey1993}
Lacey C., Cole S., 1993, MNRAS, 262, 627

\bibitem{lacey2008}
Lacey C. G., Baugh C. M., Frenk C. S., Silva L., Granato G. L., Bressan A., 2008, MNRAS, 385, 1155

\bibitem{lacey2010}
Lacey C. G., Baugh C. M., Frenk C. S., Benson A. J., Orsi A., Silva L., Granato G. L., Bressan A., 2010, MNRAS, 405, 2

\bibitem{}
Lacey C.G., Baugh C.M., Frenk C.S., Benson A.J., 2011, MNRAS, 412, 1828

\bibitem{Lagache}
Lagache G., Dole H., Puget J.-L., 2003, MNRAS, 338, 555

\bibitem{}
Lagache G., Bavouzet N., Fernandez-Conde N., Ponthieu N., 
Rodet T., Dole, H., Miville-Deschenes M.-A., Puget J.-L., 
2007, ApJ, 665, L89.


\bibitem{Le Delliou2006}
Le Delliou M., Lacey C., Baugh C. M., Morris S. L., 2006, MNRAS, 365, 712
	
\bibitem{Negrello}
Negrello M., Perrotta F., Gonzalez-Nuevo J., Silva L., de Zotti G., Granato G. L., Baccigalupi C., Danese L., 2007, MNRAS, 377, 157

\bibitem{Neistein}
Neistein E., Maccio A. V., Dekel A., 2010, MNRAS, 403, 984
 
\bibitem{}
Oliver S.J., et~al., 2011, A\&A, 518, L21 

\bibitem{orsi}
Orsi A., Lacey C. G., Baugh C. M., Infante L., 2008, MNRAS, 391, 1589

\bibitem{Papovich}
Papovich C.~et al., 2004, ApJS, 154, 70


\bibitem{}
Parkinson H., Cole S., Helly J., 2008, MNRAS, 383, 557

\bibitem{PS}
Peacock J. A., Smith R. E., 2000, MNRAS, 318, 1144

\bibitem{mission overview}
Planck Collaboration, 2011a,  2011, A\&A, 536, A1

\bibitem{CIB}
Planck Collaboration, 2011b, 2011, A\&A, 536, A18

\bibitem{LFI}
Planck Collaboration, 2011c, 2011, A\&A, 536, A5

\bibitem{HFI}
Planck Collaboration, 2011d, 2011, A\&A, 536, A6

\bibitem{HFI217}
Planck Collaboration, 2011e, 2011, A\&A,536, A13

\bibitem{Penin}
Penin A., et al. 2012a, A\&A in press, arXiv1105.1463

\bibitem{Penin2011}
Penin A., Dore O., Lagache G., Bethermin M., 2012b, A\&A, 537, A137

\bibitem{Righi}
Righi M., Hernandez-Monteagudo C., Sunyaev R. A., 2008, A\&A, 478, 685

\bibitem{Scott}
Scott S. E., Fox M. J., Dunlop J. S., Serjeant S., Peacock J. A., Ivison R. J., Oliver S., Mann R. G.~et al., 2002, MNRAS, 331,817


\bibitem{Sehgal}
Sehgal N., et al., 2010, ApJ, 709, 920

\bibitem{Shang}
Shang C., Haiman Z., Knox L., Oh S. P., 2012, MNRAS, 421, 2832

\bibitem{Sheth}
Sheth R. K., Mo H. J., Tormen G., 2001, MNRAS, 323, 1

\bibitem{Silva}
Silva L., Granato G. L., Bressan A., Danese L, 1998, ApJ, 509, 103

\bibitem{Smail}
Smail I., Ivison R. J., Blain A. W., Kneib J.-P., 2002, MNRAS, 331, 495

\bibitem{Smith}
Smith R. E., Peacock J. A., Jenkins A., White S. D. M., Frenk C. S., Pearce F. R., Thomas P. A., Efstathiou G., Couchman H. M. P., 2003, MNRAS, 341, 1311

\bibitem{Song}
Song Y. -S., Cooray A., Knox L., Zaldarriaga M., 2003, ApJ, 590, 664

\bibitem{}
Springel V., et~al. 2005, Nature, 435, 629


\bibitem{SZ}
Sunyaev R. A.,  Zel'dovich Ia. B., 1980, ARA\&A, 18, 537

\bibitem{}
Tegmark M., Efstathiou G., 1996, MNRAS, 281, 1297

\bibitem{Vega}
Vega O., Clemens M. S., Bressan A., Granato G. L., Silva L., Panuzzo P., 2008, A\&A, 484, 631

\bibitem{Vielva}
Vielva P., Mart'nez-Gonz‡lez E., Gallegos J. E., Toffolatti L., Sanz J. L., 2003, MNRAS, 344, 89

\bibitem{}
Viero M. P., et~al., 2009, ApJ, 707, 1766


\bibitem{Xia}
Xia J., Negrello M., Lapi A., de Zotti G, Danese L., Viel M., 2012, MNRAS, in press.  

\end{thebibliography}
\end{document}